\documentclass[sigconf]{acmart}

\usepackage{makecell}
\usepackage{multirow}
\usepackage{longtable}
\usepackage{booktabs}
\usepackage{threeparttable}
\usepackage{subcaption}
\usepackage{footmisc}
\usepackage{enumitem}
\usepackage{balance}
\AtBeginDocument{%
  }

\copyrightyear{2026}
\acmYear{2026}
\setcopyright{cc}
\setcctype{by}
\acmConference[KDD '26]{Proceedings of the 32nd ACM SIGKDD Conference on Knowledge Discovery and Data Mining V.1}{August 09--13, 2026}{Jeju Island, Republic of Korea}
\acmBooktitle{Proceedings of the 32nd ACM SIGKDD Conference on Knowledge Discovery and Data Mining V.1 (KDD '26), August 09--13, 2026, Jeju Island, Republic of Korea}
\acmPrice{}
\acmDOI{10.1145/3770854.3785694}
\acmISBN{979-8-4007-2258-5/2026/08}
\begin{document}

\title{ITDR: An Instruction Tuning Dataset for Enhancing Large Language Models in Recommendations}



\author{Zekun Liu}
\affiliation{%
  \institution{Beijing Jiaotong University}
\department{School of Computer Science and Technology}
  \city{Beijing}
  \country{China}}
\email{24120357@bjtu.edu.cn}

\author{Xiaowen Huang}
\authornote{Corresponding author}
\affiliation{%
  \institution{Beijing Jiaotong University}
   \department{School of Computer Science and Technology}
   \department{Beijing Key Laboratory of Traffic Data Mining and Embodied Intelligence}
   \department{Key Laboratory of Big Data \& Artificial Intelligence in Transportation, Ministry of Education}
  \city{Beijing}
  \country{China}
}
\email{xwhuang@bjtu.edu.cn}

\author{Jitao Sang}
\affiliation{%
  \institution{Beijing Jiaotong University}
   \department{School of Computer Science and Technology}
   \department{Beijing Key Laboratory of Traffic Data Mining and Embodied Intelligence}
   \department{Key Laboratory of Big Data \& Artificial Intelligence in Transportation, Ministry of Education}
  \city{Beijing}
  \country{China}
}
\email{jtsang@bjtu.edu.cn}





\renewcommand{\shortauthors}{Zekun Liu, Xiaowen Huang and Jitao Sang}

\begin{abstract}
Large language models (LLMs) have demonstrated outstanding performance in natural language processing tasks. However, in the field of recommender systems, due to the inherent structural discrepancy between user behavior data and natural language, LLMs struggle to effectively model the associations between user preferences and items. Although prompt-based methods can generate recommendation results, their inadequate understanding of recommendation tasks leads to constrained performance. To address this gap, we construct a comprehensive instruction tuning dataset, ITDR, which encompasses seven subtasks across two root tasks: user-item interaction and user-item understanding. The dataset integrates data from 13 public recommendation datasets and is built using manually crafted standardized templates, comprising approximately 200,000 instances. Experimental results demonstrate that ITDR significantly enhances the performance of mainstream open-source LLMs such as GLM-4, Qwen2.5, Qwen2.5-Instruct and LLaMA-3.2 on recommendation tasks. Furthermore, we analyze the correlations between tasks and explore the impact of task descriptions and data scale on instruction tuning effectiveness. Finally, we perform comparative experiments against closed-source LLMs with massive parameters. Our tuning dataset ITDR, the fine-tuned large recommendation models, all LoRA modules, and the complete experimental results are available at https://github.com/hellolzk/ITDR.
\end{abstract}


\maketitle
\begin{figure*}[t]
\centering
\includegraphics[width=0.75\linewidth]{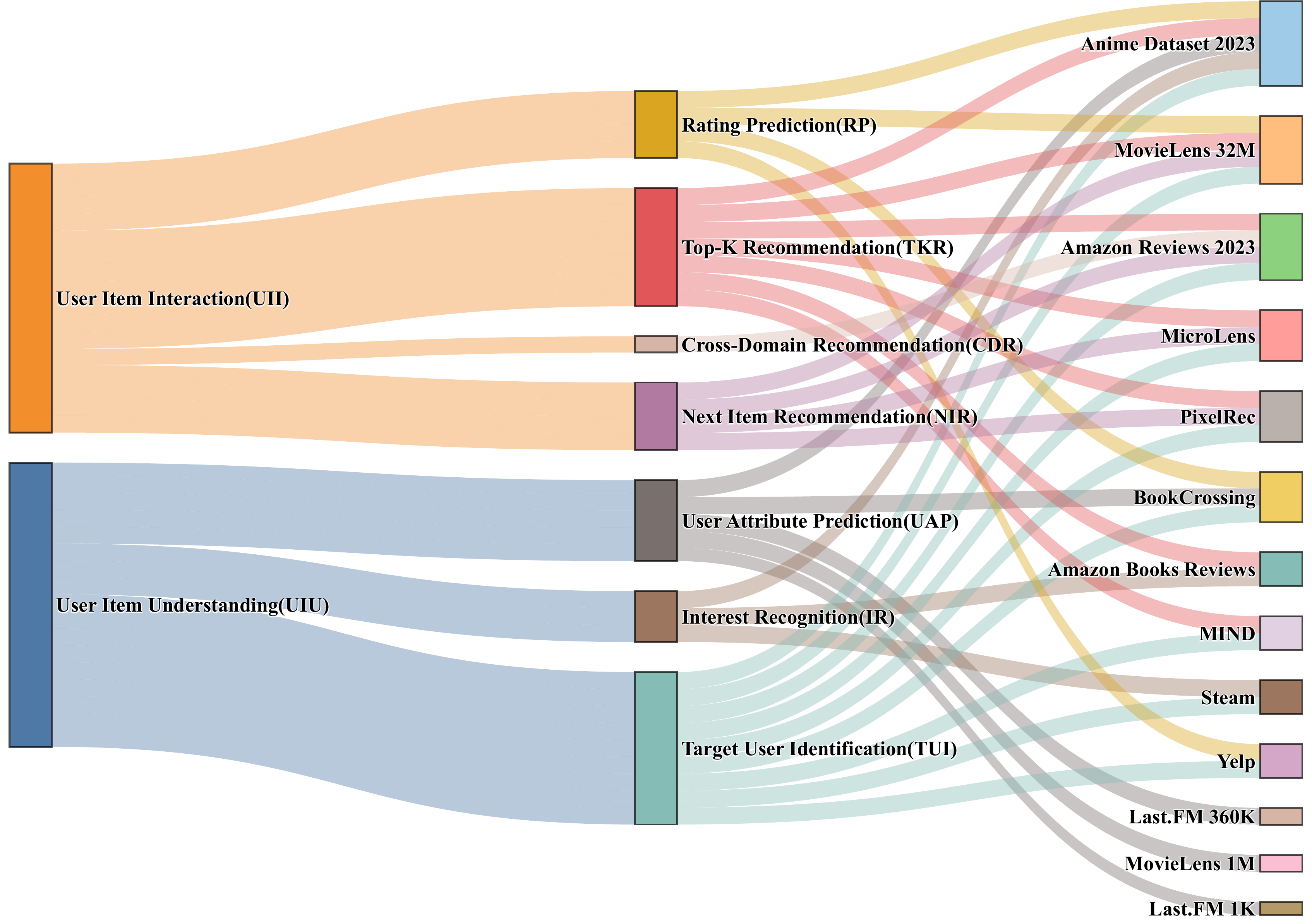} 
\caption{ITDR architecture: tasks, subtasks, and datasets.}
\label{fig:ITDR Architecture}
\end{figure*}
\section{Introduction}
Recommender systems, as a core technology to mitigate information overload, have been catalyzing innovation in personalized user experiences across various domains, such as e-commerce, social media, and content platforms~\cite{REVIEWOFRECOMMENDERSYSTEMS}. With the groundbreaking advancements in LLMs~\cite{zhao2025surveylargelanguagemodels}, effectively leveraging LLMs to enhance recommendation performance has emerged as a pivotal research focus in both academia and industry~\cite{lin2025can}. Among these advancements, instruction tuning techniques are regarded as a critical paradigm to achieve this goal~\cite{RecLM,tallrec,Recommendationasinstructionfollowing,Parameter-efficientconversationalrecommender,lin2024rella,Data-efficientFine-tuningforLLM-basedRecommendation}, by virtue of their ability to significantly improve model understanding of specific user intents and task requirements.

However, current recommendation methods that leverage instruction tuning of LLMs face a severe data bottleneck~\cite{wu2024surveylargelanguagemodels}. First, traditional recommendation datasets primarily consist of ID-based explicit or implicit user-item interaction records (e.g., ratings or clicks) and generally lack structured task descriptions and diverse natural language instructions--both of which are essential prerequisites for effective LLM fine-tuning in recommendation scenarios. Second, the heterogeneous nature of recommendation tasks necessitates that training data comprehensively cover multifaceted task scenarios and user behavior patterns. The limitations of existing datasets hinder the generalization and adaptability of LLMs in recommendation scenarios, highlighting the urgency of constructing high-quality, diverse instruction datasets.

To address these challenges, we propose ITDR--a large-scale \textbf{I}nstruction \textbf{T}uning \textbf{D}ataset specifically tailored for \textbf{R}ecommendation systems. ITDR integrates multi-dimensional data from 13 classic recommendation benchmarks (including MovieLens~\cite{MovieLens}, Amazon Reviews~\cite{AmazonReviews2023}, etc.), categorized into two core task classes--user-item interaction (\textbf{UII}) and user-item understanding (\textbf{UIU})--and further subdivided into seven distinct subtasks. Each subtask is equipped with a meticulously designed task description to guide the LLM's decision-making process. These subtasks comprehensively cover key recommendation scenarios such as rating prediction, sequential recommendation, and interest recognition, ensuring the dataset's broad applicability. ITDR ultimately comprises nearly 200,000 high-quality instructions, and we commit to publicly releasing the dataset and all fine-tuned models to foster continued progress in this field. 

Based on ITDR, we conducted systematic instruction tuning experiments on multiple open-source LLMs. Experimental results demonstrate that ITDR significantly enhances the performance of language models of varying scales across diverse recommendation tasks. 

In summary, the key contributions of this work are as follows:

1. The construction of the ITDR benchmark dataset--a large-scale instruction tuning dataset for recommender systems with comprehensive task coverage, containing nearly 200,000 high-quality instructions across two root tasks and seven subtasks.

2. Rigorous experimental validation of ITDR’s effectiveness in enhancing the recommendation performance of LLMs, with thorough ablation studies providing critical insights for future optimization and dataset expansion in LLM-based recommender systems.

\section{Related Work}
\subsection{Large Language Models for Recommendation}
LLMs have demonstrated remarkable capabilities in natural language understanding and generation, achieving successful applications in various tasks such as text summarization and question answering. Leveraging their strengths in capturing complex linguistic patterns, researchers have begun exploring the potential of LLMs in recommender systems. Current research can be broadly categorized into two main approaches.The first approach directly employs LLMs as recommender systems by designing sophisticated prompt templates to harness their zero/few-shot~\cite{gpt3} learning capabilities. Representative works include: Chat-REC~\cite{Chat-rec}, which transforms user profiles and interaction histories into prompts to enable conversational recommender systems; LLM4CDR~\cite{LLM4CRD} constructs context-aware prompts based on user's purchase histories in source domains and cross-domain shared features to enhance cross-domain recommendation performance; and RecPrompt~\cite{Recprompt} develops an adaptive prompt framework for news recommendation that automatically optimizes prompt design.The second approach integrates LLMs with traditional recommender systems to augment their capabilities. Notable examples include Llmrec~\cite{Llmrec}, which enhances graph representations of user-item interactions by incorporating LLM-generated features to alleviate data sparsity in implicit feedback scenarios, and LLM-CF~\cite{LLM-CF} distills the world knowledge and reasoning capabilities of LLMs into collaborative filtering frameworks, significantly improving recommendation effectiveness. While these methods can enhance LLMs‘ capabilities in recommendation tasks to some extent, they essentially rely on prompt engineering to elicit emergent abilities and align the models with recommendation objectives. However, due to the complexity of user behavior patterns and item features, as well as the scarcity of recommendation-related concepts in natural language text, larger models without fine-tuning do not perform well in some recommendation tasks (e.g., sequence recommendation)~\cite{Ischatgptagoodrecommender}.

Unlike existing studies, our work focuses on systematically enhancing the recommendation capabilities of LLMs through instruction tuning. By improving the model's semantic understanding of recommendation instructions and task responsiveness, we significantly enhance its practicality and adaptability in diverse recommendation scenarios.

\subsection{Fine-tuning for LLMs}
\subsubsection{Instruction tuning}
Instruction tuning is a methodology for adapting pre-trained LLMs using structured datasets expressed in natural language~\cite{Finetunedlanguagemodelsarezero-shotlearners}. This training paradigm enhances a model's ability to comprehend and execute human instructions. Conceptually, IT shares similarities with both supervised fine-tuning approaches~\cite{Traininglanguagemodelstofollowinstructionswithhumanfeedback} and multi-task learning frameworks based on prompt-based training~\cite{Multitaskpromptedtrainingenableszero-shottaskgeneralization}. The effectiveness of instruction tuning arises from its dual capacity: it not only improves performance on tasks encountered during explicit training but also fosters generalization capabilities that transfer to novel, unseen tasks~\cite{Multitaskpromptedtrainingenableszero-shottaskgeneralization}.

In terms of technical implementation, instruction tuning necessitates the reformulation of training data into input-output pairs $(x, y)$, where $x$ denotes human instructions expressed in natural language and $y$ represents the corresponding model responses. This format not only explicitly encodes the semantic descriptions of target tasks but also requires the systematic transformation of raw training samples into natural language forms aligned with human expression, thereby constructing semantically complete and logically coherent instructional contexts. Building upon this foundation, Large Language Models (LLMs) can be fine-tuned using these $(x, y)$ pairs based on the autoregressive language modeling objective:
\begin{equation}
\max _\theta \mathbb{E}_{(x, y) \sim \mathcal{D}}\left[\sum_{t=1}^{|y|} \log P_\theta\left(y_t|x, y_{<t}\right)\right].
\end{equation}

where $\theta$ represents the set of trainable parameters of the large language model (LLM), $y_t$ denotes the $t$-th token in the target sequence $y$, $y_{<t}$ refers to the sequence of tokens generated prior to time step $t$, and $\mathcal{D}$ denotes the empirical distribution over the training dataset. This objective maximizes the expected log-likelihood of generating each token autoregressively, conditioned on the input and previous outputs.

\subsubsection{Fine-tuning with Low-rank Adaptation (LoRA)}
Full fine-tuning of large language models typically demands substantial computational resources. To address this challenge, Low-Rank Adaptation~\cite{hu2021loralowrankadaptationlarge} (LoRA) has been proposed as an efficient parameter-efficient fine-tuning approach. The core idea of LoRA is to inject learnable low-rank matrices into specific layers of the Transformer architecture to approximate the weight updates required during fine-tuning of the pre-trained model. Specifically, for a pre-trained weight matrix $ {W} \in \mathbb{R}^{d_{\text{out}} \times d_{\text{in}}} $, its update $ \Delta {W} $ is factorized as the product of two low-rank matrices:
\[
\Delta {W} = {B}{A},
\]
where $ {B} \in \mathbb{R}^{d_{\text{out}} \times r} $ and $ {A} \in \mathbb{R}^{r \times d_{\text{in}}} $ are trainable upward and downward projection matrices, respectively, and the rank $ r $ is significantly smaller than both $ d_{\text{in}} $ and $ d_{\text{out}} $. This low-rank parameterization drastically reduces the number of trainable parameters and enhances adaptation efficiency.

In practice, LoRA is primarily applied to the query and value projection matrices within the multi-head attention mechanism of Transformer layers. Given an input vector $ {h} $ to such a linear transformation, the modified output $ {h}' $ under LoRA is expressed as:
\begin{equation}
{h}' = \left( {W} + \frac{\alpha}{r} \Delta {W} \right) {h} = {W}{h} + \frac{\alpha}{r} {B}{A}{h},
\end{equation}
where the original weights $ {W} $ remain frozen during fine-tuning, and $ \alpha $ is a tunable scaling factor that modulates the influence of the low-rank update relative to the base weights.

This strategy enables effective task-specific adaptation with minimal architectural modification, circumventing the need for full retraining of all model parameters while preserving strong downstream performance--all at a fraction of the computational cost.

In this work, we enhance LLM performance on recommendation tasks via instruction tuning with a carefully designed dataset that captures key characteristics and real-world scenarios of recommender systems. Experiments show the dataset effectively supports tuning and significantly boosts performance across multiple tasks, highlighting its value and applicability.


\section{ITDR: Instruction Tuning Dataset for Recommendation}
In this section, we provide a detailed introduction to ITDR. Given the substantial resources required to construct a comprehensive instruction tuning dataset that covers diverse recommendation scenarios, we adopt the methodology of prior work~\cite{zhu2024intersunlockingpowerlarge} by transforming existing recommendation datasets into an instructional format for efficient construction. Based on the characteristics of recommendation tasks, we categorize all tasks and their corresponding datasets into two root tasks: User-Item Interaction (\textbf{UII}) and User-Item Understanding (\textbf{UIU}). The complete taxonomy of tasks and their associated datasets is illustrated in Figure~\ref{fig:ITDR Architecture}. Below, we elaborate on the specific definitions of these two root tasks and their corresponding subtasks, along with the data construction process.
\subsection{Task Definition}
\subsubsection{User-Item Interaction(UII)}
In recommendation systems, user-item interactions refer to behavioral records between users and the content recommended by the system, typically including explicit or implicit signals such as clicks, views, purchases, and ratings. The core objective is to characterize the relationship between user preferences and item features, thereby providing a basis for personalized recommendations. In the recommendation process, modeling user-item interactions is a critical factor determining the accuracy of the recommendation system and the user experience. To this end, we employed four classical interaction recommendation tasks: Rating Prediction(\textbf{RP}), Top-K Recommendation(\textbf{TKR}), Cross-Domain Recommendation(\textbf{CDR}) and Next Item Recommendation(\textbf{NIR}), requiring the model to mine latent patterns in the interaction data and predict user’s interest levels in unexposed items. The following paragraphs provide detailed introductions to each subtask under the root task UII and list the source datasets we selected.
\paragraph{Rating Prediction (RP)}
Rating prediction is a classic task in recommendation systems, which aims to predict a user's potential rating for unrated items by analyzing their historical rating behaviors. We select four representative public datasets: MovieLens 32M~\cite{MovieLens}, BookCrossing~\cite{Bookcrossing}, Anime Dataset 2023\footnote{\url{https://www.kaggle.com/datasets/dbdmobile/myanimelist-dataset}\label{fn:Anime Dataset 2023}}, and Yelp~\cite{Yelp}.
\paragraph{Top-K Recommendation (TKR)}
Top-K recommendation aims to predict the top K items that a user is most likely to be interested in next by analyzing their historical interaction sequences. Unlike traditional rating prediction or click-through rate prediction, Top-K recommendation places greater emphasis on the ranking quality and diversity of the recommendation list. In this study, we utilize the following datasets:Amazon Review 2023~\cite{AmazonReviews2023}, MovieLens 32M~\cite{MovieLens}, MicroLens~\cite{MicroLens}, PixelRec~\cite{PixelRec}, Amazon Books Reviews\footnote{\url{https://www.kaggle.com/datasets/mohamedbakhet/amazon-books-reviews}\label{fn:Amazon Books Review}}, MIND~\cite{MIND}, Anime Dataset 2023\footref{fn:Anime Dataset 2023}.
\paragraph{Cross-Domain Recommendation (CDR)}
Cross-domain recommendation aims to enhance recommendation performance in a target domain by leveraging user behavior data from one or multiple source domains. Unlike single-domain recommendation, the core challenge of CDR lies in transferring user preferences and knowledge across different domains, particularly when facing data sparsity in the target domain. This study builds instruction data based on the Amazon Review 2023~\cite{AmazonReviews2023} dataset, with the specific source and target domain division scheme detailed in Table~\ref{table:cdr_task}.
   \begin{table}[h]
    \centering
    \caption{Selection strategies for source and target domains in CDR task.}
    \label{table:cdr_task}
    \begin{tabular}{cc}
        \toprule
        Source & Target \\ 
        \midrule
        Beauty  & Fashion\\ 
        CDs \& Vinyl  & Video Games \\ 
        Appliances &   Industrial \& Scientific\\ 
        Home \& Kitchen &   Grocery \& Gourmet Food \\ 
        \bottomrule
    \end{tabular}
\end{table}
\paragraph{Next Item Recommendation (NIR)}
Next item recommendation is a core task in sequential recommendation systems, with the primary objective of predicting the items a user is most likely to interact with next by modeling their historical interaction sequences (such as browsing, clicking, purchasing, etc.). We select the following publicly available datasets as data sources: Amazon Reviews 2023~\cite{AmazonReviews2023}, MicroLens~\cite{MicroLens}, MovieLens 32M~\cite{MovieLens}, and PixelRec~\cite{PixelRec}.
\subsubsection{User-Item Understanding(UIU)}
In recommendation systems, user-item understanding refers to the in-depth feature mining from the item dimension. This process involves analyzing the inherent attributes of items (such as multi-dimensional features of movies, including genre, director, cast, plot synopsis, etc.) to construct accurate item profiles, thereby achieving personalized recommendation matching for target users. We design three tasks: User Attribute Prediction(\textbf{UAP}), Interest Recognition(\textbf{IR}) and Target User Identification(\textbf{TUI}), requiring the model to deeply comprehend item features and demonstrate better adaptability in personalized recommendation scenarios. The following paragraphs provide detailed introductions to each subtask under the root task UIU and list the source datasets we selected.
\paragraph{User Attribute Prediction (UAP)}
User attribute prediction is a crucial task in building user profiles for recommendation systems. Its primary objective is to infer user’s static demographic characteristics or social attribute information by analyzing their interaction behaviors and preference patterns with items. Unlike traditional recommendation tasks, user attribute prediction does not aim to predict items that users might prefer, but rather deduces users' inherent attribute features from their behavioral data. For this subtask, we select the following datasets as data sources: Last.FM 1/360K~\cite{Last.FM1K_360K}, Anime Dataset 2023\footref{fn:Anime Dataset 2023}, BookCrossing~\cite{Bookcrossing}, and MovieLens 1M~\cite{MovieLens}.
\paragraph{Interest Recognition (IR)}
The primary objective of interest recognition is to analyze user‘s historical interaction data to identify and summarize their preferences, thereby enabling more accurate personalized services. For this subtask, we select the following datasets as data sources: Anime Dataset 2023\footref{fn:Anime Dataset 2023}, Amazon Books Reviews\footref{fn:Amazon Books Review}, and Steam~\cite{steam}.
\paragraph{Target User Identification (TUI)}
The core objective of target user identification is to analyze the attribute characteristics of items in order to identify the most likely potential user groups interested in those items. Unlike the traditional "user-to-item" recommendation logic, target user identification achieves a reverse "item-to-user" recommendation mechanism. For this subtask, we select the following datasets as data sources: Amazon Reviews 2023~\cite{AmazonReviews2023}, BookCrossing~\cite{Bookcrossing}, Anime Dataset 2023\footref{fn:Anime Dataset 2023}, Steam~\cite{steam}, Yelp~\cite{Yelp}, MicroLens~\cite{MicroLens}, PixelRec~\cite{PixelRec}, MovieLens 32M~\cite{MovieLens}, and MIND~\cite{MIND}.
\subsection{ITDR Construction}
With the task defined and dataset selected, we construct ITDR through the following three steps.

\textbf{1. Raw data collection and preprocessing. }
We manually collect various datasets related to recommender system tasks from multiple authoritative open data sources. These datasets generally exhibit high quality assurance for two key reasons: (1) they are widely adopted as benchmark datasets in the field of recommendation; (2) their data quality has been empirically validated through extensive prior research. This dual verification mechanism ensures the reliability and academic value of the collected data. After collection, we removed irrelevant item features and disqualified users.

\textbf{2. Constructing template and task descriptions.}
We design customized templates for each dataset, which are closely aligned with specific task requirements through natural language instructions. Additionally, we craft detailed task descriptions for each subtask to more effectively guide the model in accomplishing recommendation tasks.

\textbf{3. Example generation}
We use the task description as the "instruction" part of the example and the template as the "input" part of the example. For the UII task, the ground truth corresponding to each instruction can be directly obtained from the original dataset. As for the "target user identification" and "interest recognition" subtasks in the UIU task, since the original dataset does not include the corresponding true annotation data, we employ DeepSeek-V3~\cite{Deepseek-v3} to generate the reference ground truth for these tasks. Following label generation, we conduct random sampling and manual verification: specifically, for each dataset under both subtasks, we randomly sample 100 instances for human review. Our evaluation confirms that the outputs generated by DeepSeek-V3 are sufficiently accurate and fluent to serve as reliable ground-truth labels.

After completing the above steps, we construct an ITDR instruction set comprising 2 root tasks and 7 subtasks, with a total scale of 195,065 instructions. Appendix~\ref{app:statistics and experimental results} compiles comprehensive statistics and complete experimental results for ITDR. Appendix~\ref{app:Data examples} offers specific illustrative examples for each subtask. 


\subsection{Notations and Terminologies}
For clarity in describing ITDR and its associated evaluation metrics, we begin by standardizing the key notations used in this work. Table~\ref{table:notations} provides a summary of the primary symbols and their meanings.
   \begin{table}[h]
    \centering
    \caption{Key notations.}
    \label{table:notations}
    \begin{tabular}{cc}
        \toprule
        Symbol & Definition \\ 
        \midrule
        UII  & User-Item Interaction\\ 
        UIU  & User-Item Understanding \\ 
        RP &   Rating Prediction\\ 
        TKR &  Top-K Recommendation \\ 
        CDR &  Cross-Domain Recommendation \\ 
        NIR &  Next Item Recommendation  \\ 
        UAP &  User Attribute Prediction \\ 
        IR &   Interest Recognition \\ 
        TUI &  Target User Identification \\ 
        $ST$ & Set of subtasks \\
        $M$ & \makecell{Set of evaluation metrics for \\ sequential recommendation tasks} \\
        $M^{\prime}$ & Set of evaluation metrics for IR and TUI tasks\\
        $D_{st}$ & Set of all subtasks data \\
        $D_{UAP}$ & Set of UAP task\\
        $pUII$ & Performance of user-item interaction\\
        $pUIU$ & Performance of user-item understanding\\
        \bottomrule
    \end{tabular}
\end{table}
\begin{figure*}[t]
\centering
\includegraphics[width=0.80\linewidth]{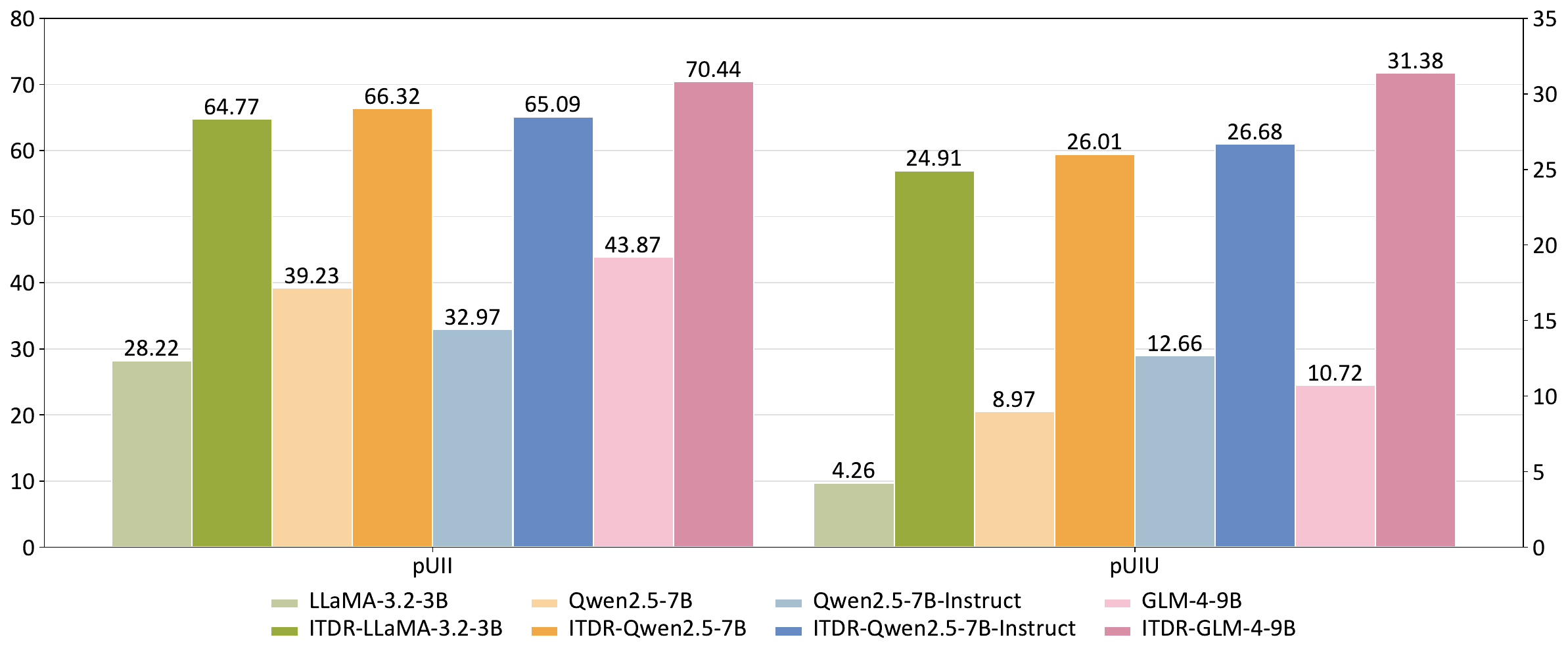} 
\caption{pUII and pUIU of all backbone models on two tasks before and after fine-tuning. "ITDR-*" refers to the backbone model "*" is fine-tuned by ITDR.}
\label{fig:Comparison_before_and_after_fine-tuning}
\end{figure*}
\section{Experiments}
In this section, we conducted extensive experiments in order to address the following key research questions:

$\bullet$ \textbf{RQ1:} Can instruction fine-tuning with ITDR enhance models' recommendation capabilities?

$\bullet$ \textbf{RQ2:} Do potential interactions exist between different root tasks?

$\bullet$ \textbf{RQ3:} What are the underlying relationships among subtasks?

$\bullet$ \textbf{RQ4:} To what extent do our designed task descriptions guide models in recommendation tasks?

$\bullet$ \textbf{RQ5:} How does data scale affect model performance?

$\bullet$ \textbf{RQ6:} What are the performance differences between fine-tuned models and closed-source LLMs on recommendation tasks?

\subsection{Backbone Models}
We select four open-source LLMs with parameter counts ranging from 3B to 9B.

$\bullet$ \textbf{LLaMA-3.2-3B}~\cite{llama3.2} is a lightweight large language model developed by Meta AI. It supports a context length of up to 128K tokens and is well-suited for natural language understanding and generation tasks in edge and mobile computing scenarios, including text summarization, multilingual dialogue, and tool invocation\footnote{\url{https://www.modelscope.cn/models/LLM-Research/Llama-3.2-3B}\label{llamafootnote}}.

$\bullet$ \textbf{Qwen2.5-7B}~\cite{qwen2.5} and its instruction-tuned variant\footnote{\url{https://www.modelscope.cn/models/Qwen/Qwen2.5-7B-Instruct}\label{qweninsfootnote}} are large language models developed by the Tongyi Qianwen team at Alibaba Cloud. These models are pretrained on a high-quality multilingual corpus exceeding 18 trillion tokens and support a context length of up to 128K tokens. They are highly capable of handling diverse tasks such as code generation, mathematical reasoning, and structured output generation, delivering robust performance in both general-purpose AI assistant and enterprise settings\footnote{\url{https://www.modelscope.cn/models/Qwen/Qwen2.5-7B}\label{qwenfootnote}}.

$\bullet$ \textbf{GLM-4-9B}~\cite{glm4} is a large language model released by Zhipu AI. It is pretrained on 10 trillion tokens of high-quality multilingual corpus, and supports a context window of 8K tokens, making it suitable for general-purpose natural language understanding and generation tasks\footnote{\url{https://www.modelscope.cn/models/ZhipuAI/glm-4-9b}\label{glmfootnote}}.

The aforementioned models represent diverse architectures, have demonstrated strong performance on standard benchmarks and are widely applied in a variety of research studies.

\subsection{Experiment Setup}

\subsubsection{Dataset Split}
ITDR comprises two root tasks with seven subtasks, totaling 195,065 data instances. In order to balance computational efficiency and fine-tuning performance, we extracted 39,061 instances and partitioned them into 31,250 training instances and 7,811 testing instances following proportional stratified sampling. All backbone models were fine-tuned and evaluated based on these data.

\subsubsection{Evaluation Metrics}
For numerical prediction tasks, including rating prediction and the prediction of user attributes such as age and birth year, we use \textbf{Root Mean Squared Error (RMSE)}~\cite{RMSE} as the evaluation metric. For other user attribute predictions, such as gender, location, and occupation, whose outputs are discrete categories, we use \textbf{Accuracy (ACC)} to evaluate the model's performance.
Our assessment of top-k recommendation and cross-domain recommendation tasks employs metrics widely adopted in prior research~\cite{NDCG}: \textbf{N@10 (Normalized Discounted Cumulative Gain at 10), H@1 (Hit Ratio at 1), and P@5 (Precision at 5)}.
In next-item prediction, we exclusively report the H@1 result, focusing on the model's ability to accurately predict the user's next interacting item.
Finally, for interest recognition and target user identification, which are inherently generative tasks, we evaluate the models' generation capabilities using \textbf{BLEU-1\&2} and \textbf{ROUGE-1\&2\&L}~\cite{bleu,rouge}. These metrics are standard for assessing the quality of generated text by comparing it against reference outputs.

Furthermore, building upon the methodology established in~\cite{zhu2024intersunlockingpowerlarge}, we define two composite metrics to holistically evaluate model performance on both user-item interaction and user-item understanding tasks. The construction methodology is as follows:
We first define the set of subtasks (ST) and the set of evaluation metrics for sequential recommendation tasks (M) as follows:
\begin{displaymath}
ST = \{RP, TKR, CDR, NIR, UAP, IR, TUI\},
\end{displaymath}
\begin{displaymath}
M = \{N@10, H@1, P@5\}.
\end{displaymath}

For each subtask, all the data it contains can be represented as the following set:
\begin{displaymath}
D_{st} = \left\{ D_{st,1}, D_{st,2}, \cdots, D_{st,n} \right\},
\end{displaymath}
where $st\in ST$, $D_{st,n}$ represents an arbitrary record in the subtask, the number of data entries contained in each subtask is $\left| D_{st} \right|$. The average performance of user-item interaction is defined as follows:
\begin{equation}
\small
\begin{aligned}
 p U I I =\frac{1}{8} & \left[\sum_{D_j \in\left\{D_{T K R}, D_{C D R}\right\}} \frac{\sum_{m \in M} \sum_{i \in D_j} m(i)}{\left|D_j\right|}\right. \\
& +\frac{\sum_{m \in M \backslash\{N @ 10, P @ 5\}} \sum_{i \in M} m(i)}{\left|D_{N I R}\right|} \\
& \left.-\lg \left(\frac{\sum_{i \in D_{R P}} R M S E(i)}{\left|D_{R P}\right|}\right)\right] \cdot 100.
\end{aligned}
\end{equation}

For the UAP subtask, since the evaluation metrics vary across different datasets, we define:
\begin{displaymath}
D_{U A P}=\left\{D_1, D_2, D_3, D_4, D_5\right\},
\end{displaymath}
where $D_1$ to $D_5$ represent Last.FM 1K, Last.FM 360K, Anime Dataset 2023, BookCrossing, and MovieLens 1M respectively.
In addition, we define the metric sets for evaluating subtasks IR and TUI as:
\begin{displaymath}
M^{\prime}=\{B L E U-1 \& 2,   R O U G E -1 \& 2 \& L\}.
\end{displaymath}

Consequently, the average performance of user-item understanding is obtained as:
\begin{equation}
\tiny
\begin{aligned}
& p U I U=\frac{1}{16}\left[\frac{\sum_{i \in\left\{D_1, D_2\right\}} \mathrm{ACC}_{\text {country }}(i)}{\left|D_1\right|+\left|D_2\right|}+\frac{\sum_{i \in\left\{D_{\mathrm{UAP}} \backslash D_4\right\}} \mathrm{ACC}_{\text {gender }}(i)}{\left|D_{\mathrm{UAP}} \backslash D_4\right|}\right. \\
& +\frac{\sum_{i \in\left\{D_3, D_4\right\}} \mathrm{ACC}_{\text {location }}(i)}{\left|D_3\right|+\left|D_4\right|}+\frac{\sum_{f \in\{\text { occupation,age group }\}} \sum_{i \in\left\{D_5\right\}} \mathrm{ACC}_f(i)}{\left|D_5\right|} \\
& -\frac{1}{4}\left(\sum_{D_k \in\left\{D_1, D_2, D_4\right\}} \lg \frac{\sum_{i \in\left\{D_k\right\}} R M S E_{\text {age }}(i)}{\left|D_k\right|}+\lg \frac{\sum_{i \in\left\{D_3\right\}} R M S E_{\text {birth year }}(i)}{\left|D_3\right|}\right) \\
& \left.+\sum_{D_l \in\left\{D_{I R}, D_{T U I}\right\}} \frac{\sum_{m \in M^{\prime}} \sum_{i \in D_l} m(i)}{\left|D_l\right|}\right] \cdot 100.
\end{aligned}
\end{equation}

Specifically, we apply a logarithmic normalization to the RMSE metric to ensure its scale aligns with that of the other metrics.
\subsubsection{Implementation Details}
To conduct our experiments, we performed LoRA fine-tuning using the LLaMA-Factory~\cite{zheng2024llamafactory} framework with BF16 mixed-precision training, setting the learning rate to 1e-4. We employed a cosine learning rate scheduler over 2 epochs, with a per-device batch size of 1 and gradient accumulation steps of 8. All experiments were executed on a cluster of 8 NVIDIA RTX 3090 24GB GPUs. For the testing phase, models were deployed and inferred using the vLLM~\cite{vllm} framework. As the primary objective of this study is to validate the effectiveness of ITDR, we excluded hyperparameter analysis from our experimental scope.
\subsection{Results and Analysis of Key Research Questions}
\begin{table*}[ht]
    \centering
    \caption{Average performance of removing different subtasks. $\downarrow$:lower is better. $\uparrow$:higher is better. “FT” means “fine-tuning". Bold: performance exceeds unfine-tuned model after subtask removal; underline: performance drops below unfine-tuned model.}
    \label{table:subtask_ablation_table}
    \begin{tabular}{lccccccccc}
        \toprule
        Subtask & w/o FT & FT & w/o RP & w/o TKR & w/o CDR & w/o NIR & w/o UAP & w/o IR & w/o TUI  \\ 
        \midrule
        RP$\downarrow$ & 0.2266  & 0.1927  & \underline{0.2459}  & 0.1734  & 0.1912  & 0.1837  & 0.1893  & 0.1818  & 0.1980   \\ 
        TKR$\uparrow$ & 0.6003  & 0.8436  & 0.8411  & \underline{0.4864}  & 0.8415  & 0.8384  & 0.8409  & 0.8389  & 0.8410   \\ 
        CDR$\uparrow$ & 0.5467  & 0.8346  & 0.8148  & 0.7963  & \textbf{0.7222}  & 0.8318  & 0.8116  & 0.8281  & 0.7970   \\ 
        NIR$\uparrow$ & 0.2956  & 0.7982  & 0.8048  & 0.8155  & 0.8053  & \textbf{0.3955}  & 0.8104  & 0.8095  & 0.8022   \\ 
        UAP$\uparrow$ & 0.0309  & 0.1529  & 0.1443  & 0.1583  & 0.1484  & 0.1450  & \textbf{0.0328}  & 0.1483  & 0.1517   \\ 
        IR$\uparrow$ & 0.1292  & 0.3756  & 0.3797  & 0.3768  & 0.3752  & 0.3772  & 0.3762  & \textbf{0.2019}  & 0.3744   \\ 
        TUI$\uparrow$ & 0.1770  & 0.4449  & 0.4449  & 0.4451  & 0.4438  & 0.4446  & 0.4459  & 0.4439  & \underline{0.0846}   \\ 
        \bottomrule
    \end{tabular}
\end{table*}
\textit{\textbf{RQ1: Can instruction fine-tuning with ITDR enhance models’ recommendation capabilities?}}

We first fine-tune the model on the entire training set and then evaluate their performance on the test set to validate the effectiveness of using ITDR for instruction tuning in recommendation tasks.

The experimental results are presented in Figure~\ref{fig:Comparison_before_and_after_fine-tuning}. Overall, after fine-tuning on ITDR, models of varying scales demonstrate significant performance improvements, confirming the effectiveness and general applicability of instruction tuning in enhancing LLM-based recommendation  capabilities. Additionally, we observe the following phenomena:

$\bullet$ Models with larger parameters show significant performance advantages. For instance, compared with LLaMA-3.2-3B, the Qwen2.5-7B series models and the GLM-4-9B perform better, indicating that larger parameters have inherent advantages in both tasks.

$\bullet$ In the UII task, GLM-4-9B exhibits optimal performance both before and after fine-tuning, which is attributed to its larger parameter count compared to other models. Notably, the experimental results show that the performance of the Qwen2.5-7B-Instruct without fine-tuning is lower than that of Qwen2.5-7B, suggesting that the instruction-tuned version may not outperform the base version in the UII task scenario.

$\bullet$ In the UIU task, the Qwen2.5-7B-Instruct without fine-tuning exhibits the best performance among the four models. This phenomenon can be attributed to the model's superior instruction-following capability, which effectively bridges the gap in parameter magnitude between it and the GLM-4-9B model. However, ITDR-GLM-4-9B achieves the optimal performance, a result that further validates the significant advantages of large-scale parametric models in terms of task learning and generalization capabilities.

$\bullet$ After fine-tuning with ITDR, the performance disparities among the four models were significantly reduced. Taking the UII task as an example, the pUII metric gap between the un-tuned LLaMA-3.2-3B and Qwen2.5-7B exceeded 10, while after fine-tuning, this gap was reduced to within 2. These results demonstrate that ITDR exhibits broad model adaptability and can effectively enhance the performance of different types of LLMs on recommendation tasks.

\noindent\textbf{\textit{RQ2: Do potential interactions exist between different root tasks?}}

\begin{figure}[h]
\centering
\subcaptionbox{Average performance of UII.\label{subfig:Task_Category_Ablation_puii}}
{\includegraphics[width=0.49\linewidth]{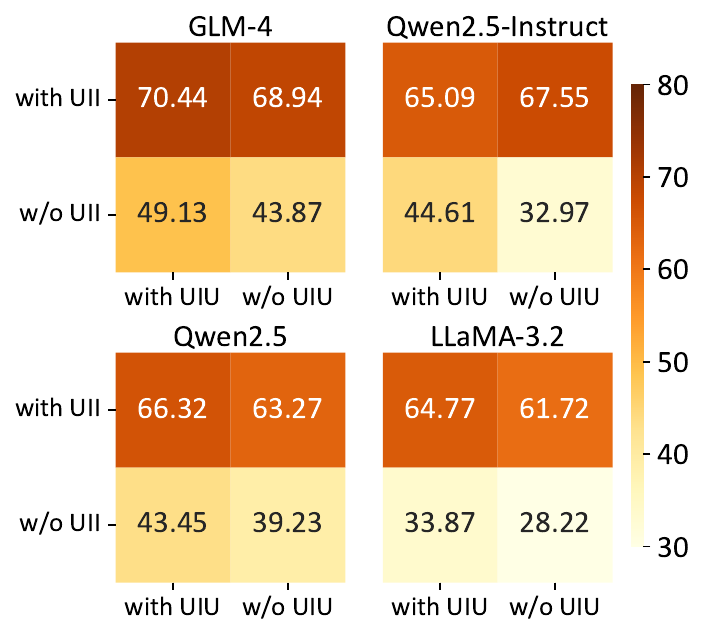}}
\hfill
\subcaptionbox{Average performance of UIU.\label{subfig:Task_Category_Ablation_puiu}}
{\includegraphics[width=0.49\linewidth]{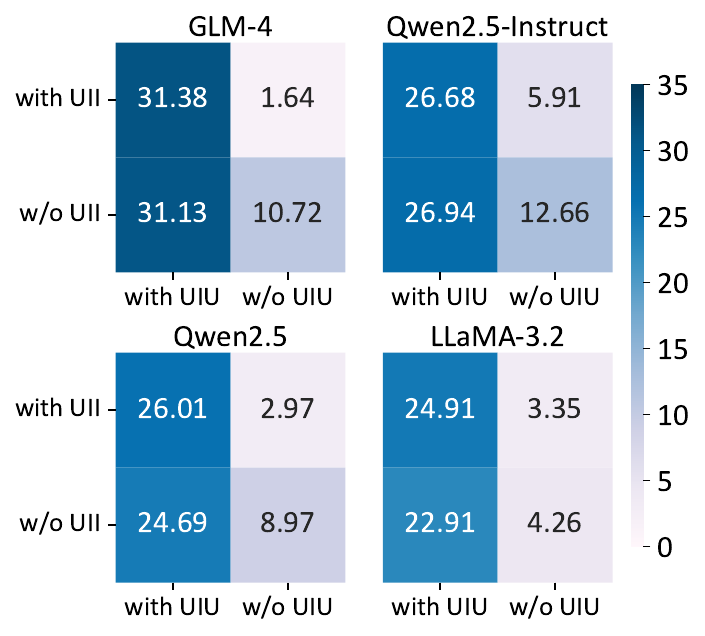}}
\caption{Average performance of removing different
root tasks.}
\label{fig:Task_Category_Ablation}
\end{figure}

In this scenario, we design category-level ablation studies to explore the correlation between the two root tasks. For each of the four models, the UIU and UII datasets are removed from the training set, and the rest of the data is retained for fine-tuning and then evaluated on the full test set (detailed performance metrics are shown in Figure~\ref{fig:Task_Category_Ablation}). The experimental design aims to reveal the potential transfer learning mechanisms and performance associations between different root tasks.

Our experimental findings can be summarized as follows:

$\bullet$ As shown in Figure~\ref{subfig:Task_Category_Ablation_puii}, models fine-tuned solely on the UIU dataset still demonstrate superior performance on the UII task compared to unfine-tuned models, a phenomenon consistently observed across all four models. This indicates that the semantic understanding capability acquired from the UIU data has produced a cross-task transfer effect on the UII task. Notably, when fine-tuned exclusively on UII data, the comprehensive performance of GLM-4-9B, Qwen2.5-7B, and LLaMA-3.2-3B models did not reach the level achieved by full-data fine-tuning, while Qwen2.5-7B-Instruct exhibited a unique reverse enhancement trend. We speculate that this discrepancy may stem from Qwen2.5-7B-Instruct's specialized instruction optimization during pre-training, which enhances its adaptability to single-task data.

$\bullet$ Figure~\ref{subfig:Task_Category_Ablation_puiu} presents the performance comparison of category-level ablation experiments on the UIU task. The result shows that for GLM-4-9B, Qwen2.5-7B, and LLaMA-3.2-3B, fine-tuning with only UIU data consistently yielded slightly inferior results compared to the full-data fine-tuning approach. However, Qwen2.5-7B-Instruct once again exhibits unique characteristics, with its single-task fine-tuning performance surpassing that of full-data training. This phenomenon further validates the heightened sensitivity of instruction-optimized models to single-task data. Particularly noteworthy is that when fine-tuned solely on UII data, all models showed significant degradation in performance on the UIU task, even falling below the level of unfine-tuned models. We attribute this performance decline to the fine-tuning process causing the model parameters to excessively favor UII task-specific feature representations, thereby compromising the general semantic understanding capabilities acquired during pre-training. The underlying mechanisms of this phenomenon remain to be further investigated.
\begin{figure}[h]
\centering
\includegraphics[width=0.8\linewidth]{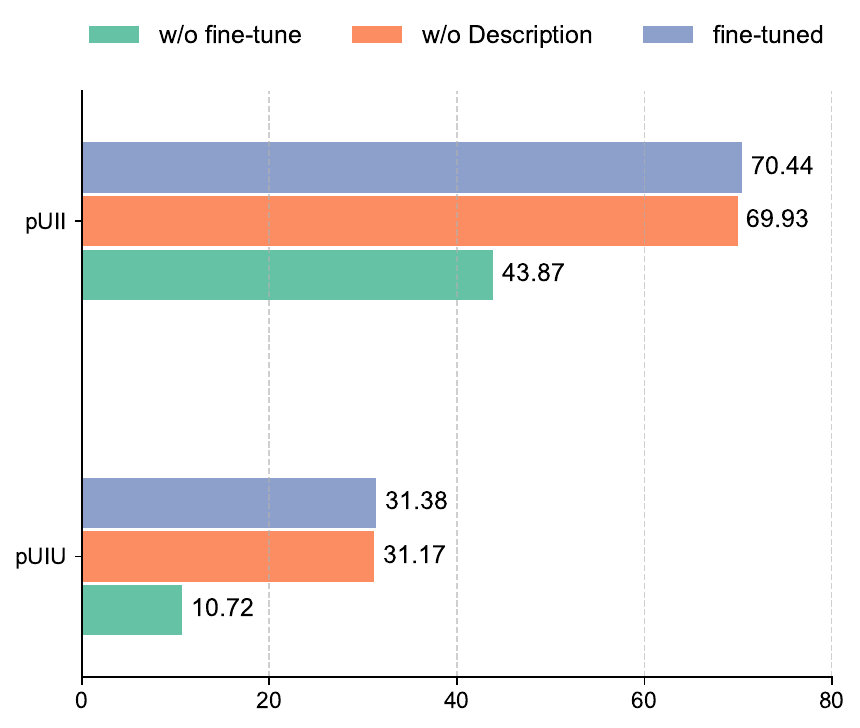} 
\caption{Ablation studies of using no task descriptions during fine-tuning.}
\label{fig:task_descriptions_fig}
\end{figure}

\noindent\textbf{\textit{RQ3: What are the underlying relationships among subtasks?}}

In this section, we investigate the synergistic mechanisms among subtasks through task-level ablation experiments. Specifically, we sequentially remove subtasks (e.g., the Top-k recommendation task) from ITDR, then fine-tune GLM-4-9B using the remaining training data, and evaluate its performance on the complete test set.

The experimental results are shown in Table~\ref{table:subtask_ablation_table}. The key experimental results are summarized below:

$\bullet$ The model demonstrates task-level generalization capabilities. For instance, after removing fine-tuning for CDR and NIR tasks, the model's performance on these tasks still surpasses that of the unfine-tuned model. We posit that the knowledge acquired through TKR task can be effectively transferred to other sequential recommendation tasks, likely because these tasks fundamentally represent different variants of sequential recommendation tasks. Similarly, the same generalization pattern is observed in IR task: despite lacking dedicated fine-tuning for IR task, the model can still leverage knowledge learned from other tasks to benefit IR task’s performance. These findings confirm the model's ability for cross-task knowledge transfer and generalization.

$\bullet$ The knowledge transfer between different subtasks does not always yield positive effects. Taking RP, TKR, and TUI tasks as examples, when their training data is excluded, the model's performance actually deteriorates compared to the unfine-tuned control group. This phenomenon not only reveals the complex interdependencies among different task datasets, but also indicates that simple task combinations may lead to negative knowledge transfer effects. These findings highlight the importance of in-depth research on data composition optimization in instruction tuning, suggesting the need to develop more rational data selection strategies to achieve optimal synergistic effects between tasks.
\noindent\textbf{\textit{RQ4: To what extent do our designed task descriptions guide models in recommendation tasks?}}

ITDR provides detailed task descriptions for each subtask, a design intended to enhance the model's task comprehension and enrich data diversity. In this section, we focus on evaluating the practical impact of these task descriptions on model performance. By removing the task description component from the instructions and fine-tuning the GLM-4-9B model, we obtained the experimental results illustrated in Figure~\ref{fig:task_descriptions_fig} and Table~\ref{table:des_ablation_table}. Based on these findings, we draw the following conclusions:
\begin{figure}[h]
\centering
\includegraphics[width=0.8\linewidth]{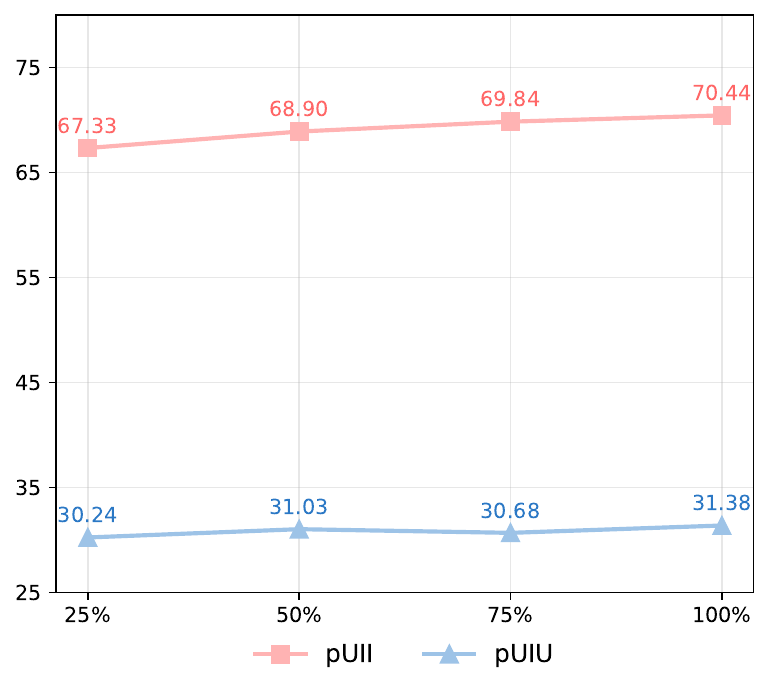} 
\caption{Average performance of using different data volumes for fine-tuning.}
\label{fig:datasize_fig}
\end{figure}
\begin{figure*}[ht]
\centering
\includegraphics[width=0.80\linewidth]{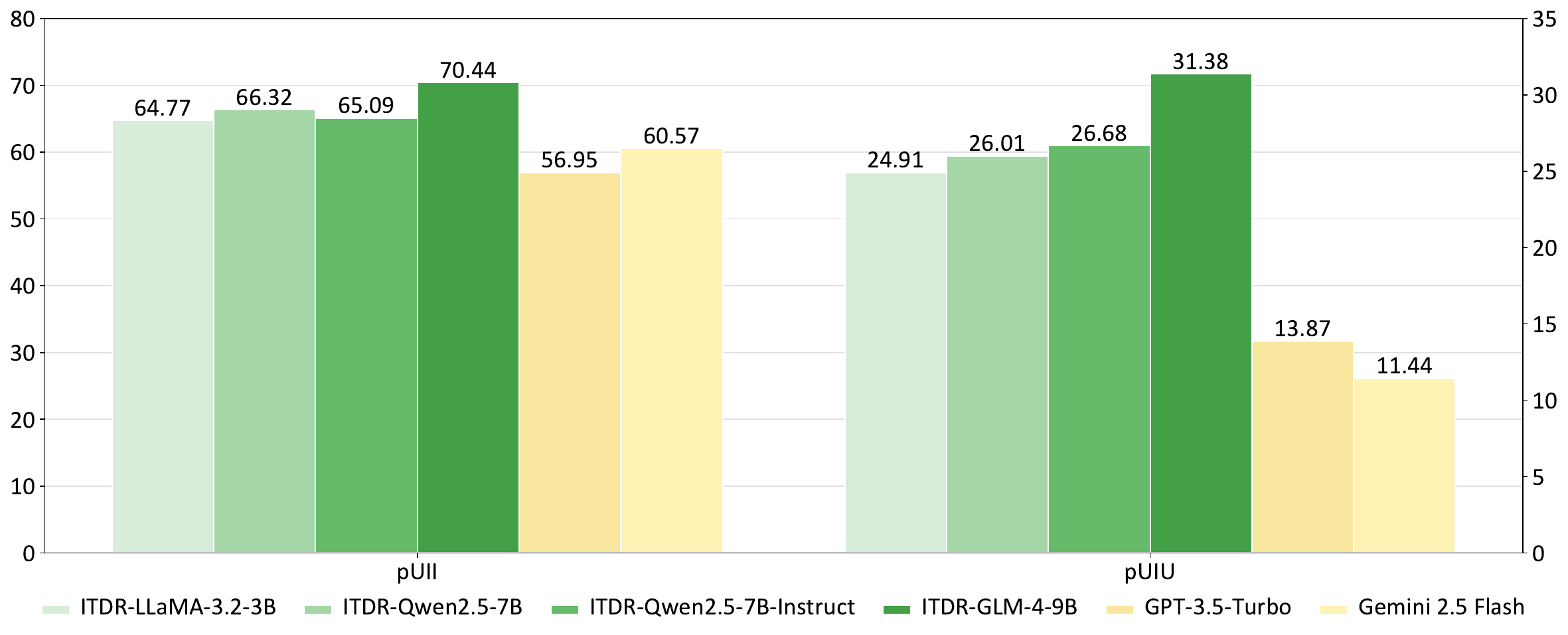} 
\caption{pUII and pUIU comparison of fine-tuned models with closed-source LLMs.}
\label{fig:Comparison_LLMs_fig}
\end{figure*}
\begin{table}[h]
    \centering
    \caption{Ablation studies of subtasks in removing task descriptions during fine-tuning. $\downarrow$:lower is better. $\uparrow$:higher is better. “FT” means “fine-tuning". The best performance is denoted in bold.}
    \label{table:des_ablation_table}
    \begin{tabular}{lccc}
        \toprule
        Subtask & w/o FT & w/o Description& FT  \\ 
        \midrule
        RP$\downarrow$  & 0.2266  & 0.2001  & \textbf{0.1927}   \\ 
        TKR$\uparrow$  & 0.6003  & 0.8411  & \textbf{0.8436}   \\ 
        CDR$\uparrow$ & 0.5467  & 0.8254  & \textbf{0.8346}   \\ 
        NIR$\uparrow$ & 0.2956  & 0.7951  & \textbf{0.7982}   \\ 
        UAP$\uparrow$ & 0.0309  & 0.1442  & \textbf{0.1529}   \\ 
        IR$\uparrow$ & 0.1292  & \textbf{0.3779}  & 0.3756   \\ 
        TUI$\uparrow$ & 0.1770  & \textbf{0.4465}  & 0.4449   \\ 
        \bottomrule
    \end{tabular}
\end{table}

$\bullet$ Task descriptions play a critical role in the model fine-tuning process. Both pUII and pUIU exhibit a declining trend when task descriptions are removed. As shown in Table~\ref{table:des_ablation_table}, the introduction of task descriptions improves model performance in most subtasks, validating the significant value of task descriptions for model training.

$\bullet$ It is noteworthy that in IR and TUI tasks, the use of task descriptions leads to a slight average performance decline. This may stem from the nature of such generative tasks: descriptive text not only fails to provide effective guidance but may also introduce additional noise that interferes with model judgment. This finding offers important insights for the optimized design of task descriptions in subsequent work, particularly when handling generative tasks, where the applicability and formulation of task descriptions require careful reconsideration.

\noindent\textbf{\textit{RQ5: How does data scale affect model performance?}}

In this section, we investigate the impact of fine-tuning data scale on model performance. Using stratified sampling, we created three subsets of training data at 25\%, 50\%, and 75\% proportions to fine-tune the GLM-4-9B. The model's performance was then evaluated on the test set, with experimental results shown in Figure~\ref{fig:datasize_fig}. Our findings reveal that:

$\bullet$ The model's performance across different tasks exhibits a positive correlation with the amount of fine-tuning data, though the sensitivity to data volume varies significantly between tasks. Specifically, the pUII metric demonstrates steady improvement as data scale increases, whereas the performance of the UIU task remains largely insensitive to variations in data quantity. This finding highlights the differential data requirements imposed by task characteristics, suggesting the need for further research on optimizing the composition and scale of instruction data based on task-specific properties to achieve more efficient model training outcomes.

\noindent\textbf{\textit{RQ6: What are the performance differences between fine-tuned models and closed-source LLMs on recommendation tasks?}}

In this section, we compared the performance differences between fine-tuned models and larger models. We selected the currently mainstream closed-source LLMs, GPT-3.5-Turbo~\cite{openai2022chatgpt} and Gemini 2.5 Flash~\cite{Gemini2025Comanici}, as benchmark models and evaluated their performance on the UII and UIU tasks. The experimental results shown in Figure~\ref{fig:Comparison_LLMs_fig} demonstrate that:

$\bullet$ The fine-tuned models demonstrate superior performance on both UII and UIU tasks compared to the baseline model with larger parameter sizes. Notably, benefiting from the advantage of massive parameter scale, larger models still exhibit significantly better performance on the UII task than untuned open-source models. Further analysis reveals distinct task preference characteristics among different models: GPT-3.5-Turbo performs more effectively on UIU tasks, while Gemini 2.5 Flash demonstrates stronger processing capabilities on UII tasks, indicating fundamental differences in how different model architectures understand and solve tasks. This series of findings reaffirms the significant value of our developed ITDR dataset in recommendation tasks.
\section{Conclusion}
In this paper, we propose ITDR, an instruction tuning dataset constructed from two dimensions--user-item interaction and user-item understanding--effectively filling the gap of natural language datasets in recommendation systems. We systematically investigate leveraging ITDR to enhance LLMs' performance on recommendation tasks. Through extensive experiments, we explore several key aspects, including task relationships, task description design, and the impact of data volume in instruction tuning, demonstrating ITDR's effectiveness in enhancing LLM-based recommendations. We expect ITDR to advance research on LLMs for recommendation and stimulate further academic exploration of instruction tuning methodologies.
\section*{Limitations and Future Work}
In this work, we construct a dataset for instruction tuning on recommendation systems and trained fine-tuned models based on this dataset. We identify the limitations of the current research and proposed directions for future improvements.

In terms of the dataset, although we have integrated seven distinct recommendation subtasks across 13 source datasets, this represents only a fraction of the available data landscape, leaving numerous potential datasets and specialized tasks untapped. Our experimental findings suggest that systematically expanding this scope to include more diverse data would not only broaden the dataset's coverage but also significantly enhance its semantic richness, thereby unlocking further gains in model performance and generalization.

Regarding model parameters, our study only validate the effectiveness of models with parameters ranging from 3B to 9B. Future work could explore the impact of instruction tuning on larger-scale models (e.g., 13B, 70B-level) and conduct comparisons with existing closed-source LLMs, which would hold significant research value.



For the evaluation methods, due to resource constraints, we primarily focused on evaluating the overall average performance across tasks. Future work could further dissect the links and transfer effects among different datasets at a finer level, thereby providing a clearer view of how data distribution characteristics affect the emerging recommendation abilities of LLMs. Such findings would offer solid guidance for optimizing data composition and sampling strategies during instruction tuning of LLMs for recommendation tasks.

\begin{acks}
This work was supported in part by the National Key Research and Development Program of China under Grant (2023YFC3310700), the National Natural Science Foundation of China (62572040, 62202041), the Beijing Natural Science Foundation (JQ24019), and the Fundamental Research Funds for the Central Universities (2025JBMC011).
\end{acks}
\bibliographystyle{ACM-Reference-Format}
\balance
\bibliography{samples/reference}

@String{Computing = "Computing" }

@String{Computer = "{IEEE} Computer" }

@String{Springer = "Springer-Verlag" }

@ArtifactSoftware{R,
    title = {R: A Language and Environment for Statistical Computing},
    author = {{R Core Team}},
    organization = {R Foundation for Statistical Computing},
    address = {Vienna, Austria},
    year = {2019},
    url = {https://www.R-project.org/},
}

@inproceedings{MIND,
    title = "{MIND}: A Large-scale Dataset for News Recommendation",
    author = "Wu, Fangzhao  and
      Qiao, Ying  and
      Chen, Jiun-Hung  and
      Wu, Chuhan  and
      Qi, Tao  and
      Lian, Jianxun  and
      Liu, Danyang  and
      Xie, Xing  and
      Gao, Jianfeng  and
      Wu, Winnie  and
      Zhou, Ming",
    editor = "Jurafsky, Dan  and
      Chai, Joyce  and
      Schluter, Natalie  and
      Tetreault, Joel",
    booktitle = "Proceedings of the 58th Annual Meeting of the Association for Computational Linguistics",
    month = jul,
    year = "2020",
    address = "Online",
    publisher = "Association for Computational Linguistics",
    url = "https://aclanthology.org/2020.acl-main.331/",
    doi = "10.18653/v1/2020.acl-main.331",
    pages = "3597--3606"
}

@article{PixelRec,
  title={An Image Dataset for Benchmarking Recommender Systems with Raw Pixels},
  author={Cheng, Yu and Pan, Yunzhu and Zhang, Jiaqi and Ni, Yongxin and Sun, Aixin and Yuan, Fajie},
  journal={arXiv preprint arXiv:2309.06789},
  year={2023}
}

@article{MovieLens,
  title={The MovieLens Datasets: History and Context},
  author={F. Maxwell Harper and Joseph A. Konstan and Joseph A.},
  journal={ACM Trans. Interact. Intell. Syst.},
  year={2016},
  volume={5},
  pages={19:1-19:19},
  url={https://api.semanticscholar.org/CorpusID:16619709}
}

@article{MicroLens,
  title={A Content-Driven Micro-Video Recommendation Dataset at Scale},
  author={Ni, Yongxin and Cheng, Yu and Liu, Xiangyan and Fu, Junchen and Li, Youhua and He, Xiangnan and Zhang, Yongfeng and Yuan, Fajie},
  journal={arXiv preprint arXiv:2309.15379},
  year={2023}
}

@book{Last.FM1K_360K,
 author = {Celma, O.},
 title = {{Music Recommendation and Discovery in the Long Tail}},
 publisher = {Springer},
 year = {2010}
}

@article{Yelp,
  title={Yelp Dataset Challenge: Review Rating Prediction},
  author={Nabiha Asghar},
  journal={ArXiv},
  year={2016},
  volume={abs/1605.05362},
  url={https://api.semanticscholar.org/CorpusID:18058702}
}

@inproceedings{steam,
  title={Self-attentive sequential recommendation},
  author={Kang, Wang-Cheng and McAuley, Julian},
  booktitle={2018 IEEE international conference on data mining (ICDM)},
  pages={197--206},
  year={2018},
  organization={IEEE}
}

@article{AmazonReviews2023,
  title={Bridging Language and Items for Retrieval and Recommendation},
  author={Hou, Yupeng and Li, Jiacheng and He, Zhankui and Yan, An and Chen, Xiusi and McAuley, Julian},
  journal={arXiv preprint arXiv:2403.03952},
  year={2024}
}

@inproceedings{Bookcrossing,
  title={Improving recommendation lists through topic diversification},
  author={Ziegler, Cai-Nicolas and McNee, Sean M and Konstan, Joseph A and Lausen, Georg},
  booktitle={Proceedings of the 14th international conference on World Wide Web},
  pages={22--32},
  year={2005}
}

@misc{zhu2024intersunlockingpowerlarge,
      title={INTERS: Unlocking the Power of Large Language Models in Search with Instruction Tuning}, 
      author={Yutao Zhu and Peitian Zhang and Chenghao Zhang and Yifei Chen and Binyu Xie and Zheng Liu and Ji-Rong Wen and Zhicheng Dou},
      year={2024},
      eprint={2401.06532},
      archivePrefix={arXiv},
      primaryClass={cs.CL},
      url={https://arxiv.org/abs/2401.06532}, 
}

@misc{Gemini2025Comanici,
    author = "Comanici, Gheorghe and many others",
    title = "Gemini 2.5: Pushing the Frontier with Advanced Reasoning, Multimodality, Long Context, and Next Generation Agentic Capabilities",
    year = "2025",
    month = "June",
    institution = "Google DeepMind",
    journal = "Technical Report"
}

@article{gpt3,
  title={Language models are few-shot learners},
  author={Brown, Tom and Mann, Benjamin and Ryder, Nick and Subbiah, Melanie and Kaplan, Jared D and Dhariwal, Prafulla and Neelakantan, Arvind and Shyam, Pranav and Sastry, Girish and Askell, Amanda and others},
  journal={Advances in neural information processing systems},
  volume={33},
  pages={1877--1901},
  year={2020}
}

@article{llama3.2,
  title={The llama 3 herd of models},
  author={Dubey, Abhimanyu and Jauhri, Abhinav and Pandey, Abhinav and Kadian, Abhishek and Al-Dahle, Ahmad and Letman, Aiesha and Mathur, Akhil and Schelten, Alan and Yang, Amy and Fan, Angela and others},
  journal={arXiv e-prints},
  pages={arXiv--2407},
  year={2024}
}

@article{glm4,
  title={Chatglm: A family of large language models from glm-130b to glm-4 all tools},
  author={GLM, Team and Zeng, Aohan and Xu, Bin and Wang, Bowen and Zhang, Chenhui and Yin, Da and Zhang, Dan and Rojas, Diego and Feng, Guanyu and Zhao, Hanlin and others},
  journal={arXiv preprint arXiv:2406.12793},
  year={2024}
}

@inproceedings{vllm,
  title={Efficient Memory Management for Large Language Model Serving with PagedAttention},
  author={Woosuk Kwon and Zhuohan Li and Siyuan Zhuang and Ying Sheng and Lianmin Zheng and Cody Hao Yu and Joseph E. Gonzalez and Hao Zhang and Ion Stoica},
  booktitle={Proceedings of the ACM SIGOPS 29th Symposium on Operating Systems Principles},
  year={2023}
}

@article{RMSE,
  author = {Chai, T. and Draxler, R. R.},
  title = {Root mean square error (RMSE) or mean absolute error (MAE)?—Arguments against avoiding RMSE in the literature},
  journal = {Geoscientific Model Development},
  volume = {7},
  number = {3},
  pages = {1247--1250},
  year = {2014},
  publisher = {Copernicus GmbH},
  doi = {10.5194/gmd-7-1247-2014},
  url = {https://gmd.copernicus.org/articles/7/1247/2014/}
}

@inproceedings{NDCG, series={WWW ’25},
   title={Uncovering Cross-Domain Recommendation Ability of Large Language Models},
   url={http://dx.doi.org/10.1145/3701716.3717850},
   DOI={10.1145/3701716.3717850},
   booktitle={Companion Proceedings of the ACM on Web Conference 2025},
   publisher={ACM},
   author={Liu, Xinyi and Wang, Ruijie and Sun, Dachun and Hakkani Tur, Dilek and Abdelzaher, Tarek},
   year={2025},
   month=may, pages={2736–2743},
   collection={WWW ’25} }

@inproceedings{bleu,
    title = "{B}leu: a Method for Automatic Evaluation of Machine Translation",
    author = "Papineni, Kishore  and
      Roukos, Salim  and
      Ward, Todd  and
      Zhu, Wei-Jing",
    editor = "Isabelle, Pierre  and
      Charniak, Eugene  and
      Lin, Dekang",
    booktitle = "Proceedings of the 40th Annual Meeting of the Association for Computational Linguistics",
    month = jul,
    year = "2002",
    address = "Philadelphia, Pennsylvania, USA",
    publisher = "Association for Computational Linguistics",
    url = "https://aclanthology.org/P02-1040/",
    doi = "10.3115/1073083.1073135",
    pages = "311--318"
}

@inproceedings{rouge,
    title = "{ROUGE}: A Package for Automatic Evaluation of Summaries",
    author = "Lin, Chin-Yew",
    booktitle = "Text Summarization Branches Out",
    month = jul,
    year = "2004",
    address = "Barcelona, Spain",
    publisher = "Association for Computational Linguistics",
    url = "https://aclanthology.org/W04-1013/",
    pages = "74--81"
}

@misc{openai2022chatgpt,
  author = {OpenAI},
  title = {Introducing {ChatGPT}},
  year = {2022},
  url = {https://openai.com/blog/chatgpt},
}

@misc{zhao2025surveylargelanguagemodels,
      title={A Survey of Large Language Models}, 
      author={Wayne Xin Zhao and Kun Zhou and Junyi Li and Tianyi Tang and Xiaolei Wang and Yupeng Hou and Yingqian Min and Beichen Zhang and Junjie Zhang and Zican Dong and Yifan Du and Chen Yang and Yushuo Chen and Zhipeng Chen and Jinhao Jiang and Ruiyang Ren and Yifan Li and Xinyu Tang and Zikang Liu and Peiyu Liu and Jian-Yun Nie and Ji-Rong Wen},
      year={2025},
      eprint={2303.18223},
      archivePrefix={arXiv},
      primaryClass={cs.CL},
      url={https://arxiv.org/abs/2303.18223}, 
}

@article{Recommendationasinstructionfollowing,
  title={Recommendation as instruction following: A large language model empowered recommendation approach},
  author={Zhang, Junjie and Xie, Ruobing and Hou, Yupeng and Zhao, Xin and Lin, Leyu and Wen, Ji-Rong},
  journal={ACM Transactions on Information Systems},
  volume={43},
  number={5},
  pages={1--37},
  year={2025},
  publisher={ACM New York, NY}
}

@article{Parameter-efficientconversationalrecommender,
  title={Parameter-efficient conversational recommender system as a language processing task},
  author={Ravaut, Mathieu and Zhang, Hao and Xu, Lu and Sun, Aixin and Liu, Yong},
  journal={arXiv preprint arXiv:2401.14194},
  year={2024}
}

@misc{wu2024surveylargelanguagemodels,
      title={A Survey on Large Language Models for Recommendation}, 
      author={Likang Wu and Zhi Zheng and Zhaopeng Qiu and Hao Wang and Hongchao Gu and Tingjia Shen and Chuan Qin and Chen Zhu and Hengshu Zhu and Qi Liu and Hui Xiong and Enhong Chen},
      year={2024},
      eprint={2305.19860},
      archivePrefix={arXiv},
      primaryClass={cs.IR},
      url={https://arxiv.org/abs/2305.19860}, 
}

@article{Chat-rec,
  title={Chat-rec: Towards interactive and explainable llms-augmented recommender system},
  author={Gao, Yunfan and Sheng, Tao and Xiang, Youlin and Xiong, Yun and Wang, Haofen and Zhang, Jiawei},
  journal={arXiv preprint arXiv:2303.14524},
  year={2023}
}

@inproceedings{LLM4CRD,
  title={Uncovering cross-domain recommendation ability of large language models},
  author={Liu, Xinyi and Wang, Ruijie and Sun, Dachun and Hakkani Tur, Dilek and Abdelzaher, Tarek},
  booktitle={Companion Proceedings of the ACM on Web Conference 2025},
  pages={2736--2743},
  year={2025}
}

@inproceedings{Recprompt,
  title={Recprompt: A self-tuning prompting framework for news recommendation using large language models},
  author={Liu, Dairui and Yang, Boming and Du, Honghui and Greene, Derek and Hurley, Neil and Lawlor, Aonghus and Dong, Ruihai and Li, Irene},
  booktitle={Proceedings of the 33rd ACM International Conference on Information and Knowledge Management},
  pages={3902--3906},
  year={2024}
}

@inproceedings{Llmrec,
  title={Llmrec: Large language models with graph augmentation for recommendation},
  author={Wei, Wei and Ren, Xubin and Tang, Jiabin and Wang, Qinyong and Su, Lixin and Cheng, Suqi and Wang, Junfeng and Yin, Dawei and Huang, Chao},
  booktitle={Proceedings of the 17th ACM international conference on web search and data mining},
  pages={806--815},
  year={2024}
}

@inproceedings{LLM-CF,
  title={Large language models enhanced collaborative filtering},
  author={Sun, Zhongxiang and Si, Zihua and Zang, Xiaoxue and Zheng, Kai and Song, Yang and Zhang, Xiao and Xu, Jun},
  booktitle={Proceedings of the 33rd ACM International Conference on Information and Knowledge Management},
  pages={2178--2188},
  year={2024}
}

@article{Ischatgptagoodrecommender,
  title={Is chatgpt a good recommender? a preliminary study},
  author={Liu, Junling and Liu, Chao and Zhou, Peilin and Lv, Renjie and Zhou, Kang and Zhang, Yan},
  journal={arXiv preprint arXiv:2304.10149},
  year={2023}
}

@article{Finetunedlanguagemodelsarezero-shotlearners,
  title={Finetuned language models are zero-shot learners},
  author={Wei, Jason and Bosma, Maarten and Zhao, Vincent Y and Guu, Kelvin and Yu, Adams Wei and Lester, Brian and Du, Nan and Dai, Andrew M and Le, Quoc V},
  journal={arXiv preprint arXiv:2109.01652},
  year={2021}
}

@article{Traininglanguagemodelstofollowinstructionswithhumanfeedback,
  title={Training language models to follow instructions with human feedback},
  author={Ouyang, Long and Wu, Jeffrey and Jiang, Xu and Almeida, Diogo and Wainwright, Carroll and Mishkin, Pamela and Zhang, Chong and Agarwal, Sandhini and Slama, Katarina and Ray, Alex and others},
  journal={Advances in neural information processing systems},
  volume={35},
  pages={27730--27744},
  year={2022}
}

@article{Multitaskpromptedtrainingenableszero-shottaskgeneralization,
  title={Multitask prompted training enables zero-shot task generalization},
  author={Sanh, Victor and Webson, Albert and Raffel, Colin and Bach, Stephen H and Sutawika, Lintang and Alyafeai, Zaid and Chaffin, Antoine and Stiegler, Arnaud and Scao, Teven Le and Raja, Arun and others},
  journal={arXiv preprint arXiv:2110.08207},
  year={2021}
}

@article{Deepseek-v3,
  title={Deepseek-v3 technical report},
  author={Liu, Aixin and Feng, Bei and Xue, Bing and Wang, Bingxuan and Wu, Bochao and Lu, Chengda and Zhao, Chenggang and Deng, Chengqi and Zhang, Chenyu and Ruan, Chong and others},
  journal={arXiv preprint arXiv:2412.19437},
  year={2024}
}

@misc{hu2021loralowrankadaptationlarge,
      title={LoRA: Low-Rank Adaptation of Large Language Models}, 
      author={Edward J. Hu and Yelong Shen and Phillip Wallis and Zeyuan Allen-Zhu and Yuanzhi Li and Shean Wang and Lu Wang and Weizhu Chen},
      year={2021},
      eprint={2106.09685},
      archivePrefix={arXiv},
      primaryClass={cs.CL},
      url={https://arxiv.org/abs/2106.09685}, 
}

@misc{RecLM,
      title={RecLM: Recommendation Instruction Tuning}, 
      author={Yangqin Jiang and Yuhao Yang and Lianghao Xia and Da Luo and Kangyi Lin and Chao Huang},
      year={2025},
      eprint={2412.19302},
      archivePrefix={arXiv},
      primaryClass={cs.IR},
      url={https://arxiv.org/abs/2412.19302}, 
}

@misc{Data-efficientFine-tuningforLLM-basedRecommendation,
      title={Data-efficient Fine-tuning for LLM-based Recommendation}, 
      author={Xinyu Lin and Wenjie Wang and Yongqi Li and Shuo Yang and Fuli Feng and Yinwei Wei and Tat-Seng Chua},
      year={2024},
      eprint={2401.17197},
      archivePrefix={arXiv},
      primaryClass={cs.IR},
      url={https://arxiv.org/abs/2401.17197}, 
}

@misc{lin2024rella,
      title={ReLLa: Retrieval-enhanced Large Language Models for Lifelong Sequential Behavior Comprehension in Recommendation}, 
      author={Jianghao Lin and Rong Shan and Chenxu Zhu and Kounianhua Du and Bo Chen and Shigang Quan and Ruiming Tang and Yong Yu and Weinan Zhang},
      year={2024},
      eprint={2308.11131},
      archivePrefix={arXiv},
      primaryClass={cs.IR},
      url={https://arxiv.org/abs/2308.11131}, 
}

@misc{qwen2.5,
      title={Qwen2.5 Technical Report}, 
      author={Qwen and : and An Yang and Baosong Yang and Beichen Zhang and Binyuan Hui and Bo Zheng and Bowen Yu and Chengyuan Li and Dayiheng Liu and Fei Huang and Haoran Wei and Huan Lin and Jian Yang and Jianhong Tu and Jianwei Zhang and Jianxin Yang and Jiaxi Yang and Jingren Zhou and Junyang Lin and Kai Dang and Keming Lu and Keqin Bao and Kexin Yang and Le Yu and Mei Li and Mingfeng Xue and Pei Zhang and Qin Zhu and Rui Men and Runji Lin and Tianhao Li and Tianyi Tang and Tingyu Xia and Xingzhang Ren and Xuancheng Ren and Yang Fan and Yang Su and Yichang Zhang and Yu Wan and Yuqiong Liu and Zeyu Cui and Zhenru Zhang and Zihan Qiu},
      year={2025},
      eprint={2412.15115},
      archivePrefix={arXiv},
      primaryClass={cs.CL},
      url={https://arxiv.org/abs/2412.15115}, 
}

@inproceedings{tallrec,
  title={Tallrec: An effective and efficient tuning framework to align large language model with recommendation},
  author={Bao, Keqin and Zhang, Jizhi and Zhang, Yang and Wang, Wenjie and Feng, Fuli and He, Xiangnan},
  booktitle={Proceedings of the 17th ACM conference on recommender systems},
  pages={1007--1014},
  year={2023}
}

@misc{zheng2024llamafactory,
      title={LlamaFactory: Unified Efficient Fine-Tuning of 100+ Language Models}, 
      author={Yaowei Zheng and Richong Zhang and Junhao Zhang and Yanhan Ye and Zheyan Luo and Zhangchi Feng and Yongqiang Ma},
      year={2024},
      eprint={2403.13372},
      archivePrefix={arXiv},
      primaryClass={cs.CL},
      url={https://arxiv.org/abs/2403.13372}, 
}

@article{lin2025can,
  title={How can recommender systems benefit from large language models: A survey},
  author={Lin, Jianghao and Dai, Xinyi and Xi, Yunjia and Liu, Weiwen and Chen, Bo and Zhang, Hao and Liu, Yong and Wu, Chuhan and Li, Xiangyang and Zhu, Chenxu and others},
  journal={ACM Transactions on Information Systems},
  volume={43},
  number={2},
  pages={1--47},
  year={2025},
  publisher={ACM New York, NY}
}

@article{REVIEWOFRECOMMENDERSYSTEMS,
  title={A comprehensive review of recommender systems: Transitioning from theory to practice},
  author={Raza, Shaina and Rahman, Mizanur and Kamawal, Safiullah and Toroghi, Armin and Raval, Ananya and Navah, Farshad and Kazemeini, Amirmohammad},
  journal={Computer Science Review},
  volume={59},
  pages={100849},
  year={2026},
  publisher={Elsevier}
}
\appendix


\section{Statistics of ITDR and Additional Results}
\label{app:statistics and experimental results}
Detailed statistics of ITDR and experimental results are presented in Table~\ref{tab:statistics} -- Table~\ref{tab:user_item_understanding_gpt_gemini_continue}.

Table~\ref{tab:statistics} summarizes the complete statistical information of ITDR, encompassing the following dimensions: subtasks, data sources, evaluation metrics, sample size, and the average input/output tokens calculated based on the DeepSeek-v3 tokenizer.

According to the experimental data in Table~\ref{tab:user_item_interaction_old} -- Table~\ref{tab:user_item_understanding_continue_old}, the fine-tuned models show significant improvements in performance metrics across all tasks, particularly in sequence recommendation tasks (TKR, CDR, NIR) and generation tasks (IR, TUI). Taking the performance of GLM-4 before and after fine-tuning as an example, on the MicroLens dataset under the TKR task, the H@1 metric increases from 0.5520 to 0.6960; on the PixelRec dataset for the TUI task, the ROUGE-1 score also rises from 0.3359 to 0.5246. These findings strongly validate the effectiveness of ITDR in enhancing the recommendation capabilities of LLMs. 

Based on the experimental results of the ablation studies at the root tasks shown in Table~\ref{tab:user_item_interaction_categories} -- Table~\ref{tab:user_item_understanding_categories_continue}, we can draw the following conclusions: first, the UIU tasks enhance the performance of the UII tasks, which aligns with our expectations. This indicates that the knowledge learned by the LLMs through the UIU tasks can be effectively transferred and guide the reasoning process of the UII tasks. However, the study also revealed a noteworthy phenomenon: when the model is fine-tuned using only the UII data, the performance of the UIU tasks exhibits a clear decline. This counterintuitive observation suggests that there may be complex interaction mechanisms between different task datasets, requiring further research to explore their underlying relationships. At the same time, this finding provides important empirical evidence and research directions for optimizing data composition strategies in future model fine-tuning processes.

Table~\ref{tab:user_item_interaction_Ablation} -- Table~\ref{tab:user_item_understanding_Ablation2} present the experimental results of GLM-4-9B in subtask level ablation studies. We observe interaction effects among different subtasks. Specifically, when the CDR task is removed during fine-tuning, the model's performance on this task surpassed that of the unfine-tuned baseline results (as shown in Table~\ref{tab:user_item_interaction_old}). This indicates that the knowledge learned from other subtasks can be effectively transferred to this task, positively enhancing its reasoning capabilities. However, such interaction effects are not always beneficial. For instance, when the RP task is removed during fine-tuning, the model's average performance on this task is lower than the baseline level. These findings highlight the importance of formulating rational data selection strategies during the model fine-tuning process.

Based on the experimental results from Table~\ref{tab:user_item_interaction_gpt_gemini} -- Table~\ref{tab:user_item_understanding_gpt_gemini_continue}, GPT-3.5-Turbo and Gemini 2.5 Flash exhibit the following characteristics in UII and UIU tasks: First, owing to their substantial model parameters, these two LLMs demonstrate significantly superior performance in UII tasks compared to untuned small-scale models, though they still fall short of the level achieved by small models that have undergone instruction tuning. Second, in UIU tasks, their performance shows no significant difference when compared to untuned small-scale models. Notably, the two models exhibit performance divergence across different tasks: GPT-3.5-Turbo performs more prominently in UIU tasks, while Gemini 2.5 Flash shows a relative advantage in UII tasks. This phenomenon suggests that LLMs with different architectures may possess distinct strengths in specific task domains.

\section{Data Examples}
\label{app:Data examples}
Data examples for each subtask are presented in Table~\ref{tab:RP_EXP} -- Table~\ref{tab:TUI_EXP}.

Taking the instruction data from the Anime Dataset 2023 shown in Table~\ref{tab:RP_EXP} as an example, we adopt the following method to construct instruction data for the RP task: First, valid user samples with no fewer than 11 rating records are filtered. Then, the viewing records of each user are sorted in ascending chronological order. Next, the first 10 rating records are extracted as the model's prior knowledge input. Finally, the 11th rating record is used as the prediction target, with its true rating value serving as the ground truth for model performance evaluation.

Table~\ref{tab:TKR_EXP} presents the TKR task instruction construction method based on the PixelRec data source. We first filter user samples with no fewer than 11 interaction records and sort their interaction sequences in chronological order. When constructing instructions, the first 5 interaction records of each user are used as prior knowledge input for the model. Meanwhile, the subsequent 5 positive samples are mixed with 5 randomly sampled negative samples and shuffled to form a candidate set to be ranked. The output content is the correct ranking order of the samples to be sorted.


When constructing instructions for the CDR task based on the Amazon Review 2023 dataset, we first use the user’s earliest 10 interaction records from the source domain (sorted chronologically in ascending order) as prior knowledge input. Second, we select the earliest 5 positive samples from the target domain (also sorted chronologically) and combine them with 5 randomly sampled negative samples to form a candidate ranking set. The model is expected to output the correct ranking order of these samples. See Table~\ref{tab:CDR_EXP} for details.

Table~\ref{tab:NIR_EXP} presents an instruction example for the NIR task, which follows a design approach similar to the TKR task. We first filter user samples with no fewer than 6 interaction records and sort their interaction sequences in chronological order. The first 5 interaction records of each user serve as prior knowledge input for the model, while the remaining one positive sample is mixed with 4 randomly sampled negative samples and shuffled to form the candidate set for prediction. The model's objective is to accurately identify the positive sample from the candidate set.

Taking the MovieLens 1M dataset as an example, we designed an instruction data construction method for the UAP task: First, we filtered out each user's movie-watching records with ratings above 3 as positive samples and sorted them in ascending order by viewing time. Then, we used the user's static attribute features as the target variables for the model to predict. As shown in Table~\ref{tab:UAP_EXP}, we provide an example to illustrate the data construction process.

Table~\ref{tab:IR_EXP} presents an example of instruction construction for IR tasks, which is built based on the Amazon Books Reviews dataset. We first categorize user ratings into "like" (greater than 3 points) and "dislike" (less than 3 points) using a threshold of 3, then select 3 representative samples from each category, including book titles and their corresponding reviews, as prior knowledge for the model. Since this is a generative task and the original dataset lacks prediction targets, we employ the DeepSeek-V3 to generate reliable ground truth data as an evaluation benchmark, thereby providing high-quality data support for model training and evaluation.

Table~\ref{tab:TUI_EXP} demonstrates an example for constructing TUI task instructions based on the Steam dataset. First, for each game, extract its key attributes, including feature information such as the game's name, genre, tags, and gameplay. Second, randomly sample real player reviews of the game (which may contain negative feedback). Similar to IR tasks, we use DeepSeek-V3 to generate standard answer data as the expected output for the instructions.

\begin{table*}[t]
    \centering

    \begin{threeparttable}
    \caption{The statistics of all datasets.}
    \label{tab:statistics}
\begin{tabular}{llccccccc}
    \toprule
        \textbf{Task} & \textbf{Dataset} & \textbf{Metrics} &\textbf{Examples} & \textbf{Avg In} & \textbf{Avg Out} \\
    \midrule
Rating Prediction & MovieLens 32M           & \multirow{4}{*}{RMSE} & 6,250  & 162.32 & 3.00 \\
Rating Prediction                                   & BookCrossing            &                       & 4,623  & 78.47  & 1.00 \\
 Rating Prediction                                  & Anime Dataset 2023      &                       & 6,250  & 120.02 & 1.00 \\
 Rating Prediction                                  & Yelp                    &                       & 5,707  & 323.98 & 1.00 \\ \\
        Top-K Recommendation & Amazon Reviews 2023 & \multirow{7}{*}{\makecell[c]{N@10 \\ H@1 \\ P@5}} & 5,276  & 435.32 & 264.53 \\
        Top-K Recommendation                                              & MovieLens 32M       &                                            & 6,250  & 208.35 & 109.13 \\
\bottomrule
    \end{tabular}
    \begin{tablenotes}
        \item \footnotesize\textsuperscript{*}The average input/output tokens are calculated using the DeepSeek-v3 tokenizer.
    \end{tablenotes}
    \end{threeparttable}
\end{table*}

\begin{table*}[t]
    \centering
    \setlength{\tabcolsep}{1.2mm}{
    \begin{threeparttable}
    \caption{Performance on the user-item interaction tasks.}
    \label{tab:user_item_interaction_old}
    \begin{tabular}{llccccccccccc}
    \toprule
    &  & \multicolumn{5}{c}{GLM-4} & \multicolumn{2}{c}{Qwen2.5-Ins} & \multicolumn{2}{c}{Qwen2.5} & \multicolumn{2}{c}{LLaMA-3.2} \\
    \cmidrule(lr){3-7}\cmidrule(lr){8-9}\cmidrule(lr){10-11}\cmidrule(lr){12-13}
    Task \& Dataset & Metric & No FT & FT & 25\% & 50\% & 75\%& No FT & FT& No FT & FT & No FT & FT\\
    \midrule
    \multicolumn{13}{l}{\textbf{{Rating Prediction}}} \\
    MovieLens 32M 
     & RMSE & 1.1469 & 1.0339 & 0.9945 & 1.0780 & 1.0010 & 1.1151 & 1.0276 & 1.0108 & 1.0877 & 1.1239 & 1.1050  \\
    BookCrossing 
     & RMSE & 1.8562 & 1.7430 & 1.7289 & 1.7477 & 1.7242 & 1.9004 & 1.8518 & 1.9546& 1.7754 & 1.9151 & 2.0589  \\
    Anime Dataset 2023 
     & RMSE & 1.5969 & 1.4213 & 1.5179 & 1.4085 & 1.3813 & 1.5356 & 1.6050 & 1.5526& 1.5466 & 1.6952 & 1.5543  \\
    Yelp
     & RMSE & 2.1065 & 2.0196 & 1.9206 & 1.9779 & 2.0022 & 2.3952 & 2.0562 & 1.9217& 2.0011 & 2.2891 & 2.4324  \\
    \midrule
    \multicolumn{13}{l}{\textbf{{Top-K Recommendation}}} \\
\end{tabular}
     \begin{tablenotes}
        \item \footnotesize\textsuperscript{*}“FT” means “fine-tuning”. “25\%” indicates that the model was finetuned using 25\% of the training data.
    \end{tablenotes}
    \end{threeparttable}
    }
\end{table*}

\begin{table*}[t]
    \centering
    \small
    \setlength{\tabcolsep}{0.6mm}{
        \begin{threeparttable}
    \caption{Performance on the user-item understanding tasks.}
    \label{tab:user_item_understanding_old}
    \begin{tabular}{llccccccccccc}
    \toprule
    &  & \multicolumn{5}{c}{GLM-4			
} & \multicolumn{2}{c}{Qwen2.5-Ins			
} & \multicolumn{2}{c}{Qwen2.5			
} & \multicolumn{2}{c}{LLaMA-3.2		
} \\
    \cmidrule(lr){3-7}\cmidrule(lr){8-9}\cmidrule(lr){10-11}\cmidrule(lr){12-13}
    Task \& Dataset & Metric & No FT & FT & 25\% & 50\% & 75\%  & No FT & FT & No FT & FT & No FT & FT\\
    \midrule
    \multicolumn{13}{l}{\textbf{{User Attribute Prediction}}} \\
Last.FM 1K 
 & ACC(c) & 0.1892 & 0.4865 & 0.5676 & 0.5135 & 0.5405 & 0.2162 & 0.4865 & 0.1622 & 0.5676 & 0.1622 & 0.3514  \\
 & ACC(g) & 0.2702 & 0.5405 & 0.5946 & 0.6216 & 0.5946 & 0.6216 & 0.6757 & 0.2432 & 0.4054 & 0.1081 & 0.2432  \\
 & RMSE(a) & 6.3650 & 6.1489 & 6.5826 & 6.3622 & 6.2469 & 5.9780 & 5.7680 & 6.3690 & 6.1056 & 5.9512 & 6.0075  \\
Last.FM 360K 
 & ACC(c) & 0.2560 & 0.4720 & 0.4560 & 0.4400 & 0.4480 & 0.2440 & 0.3840 & 0.1480 & 0.3880 & 0.0760 & 0.4120  \\
 & ACC(g) & 0.5480 & 0.7240 & 0.7280 & 0.7440 & 0.7200 & 0.7080 & 0.7320 & 0.5320 & 0.6960 & 0.5800 & 0.7160  \\
\bottomrule
    \end{tabular}
    \begin{tablenotes}
        \item \footnotesize\textsuperscript{*}“FT” means “fine-tuning”. “25\%” indicates that the model was finetuned using 25\% of the training data.
    \end{tablenotes}
    \end{threeparttable}
    }
\end{table*}

\begin{table*}[t]
    \centering
        \caption{Continued from Table~\ref{tab:user_item_understanding_old}.}
    \label{tab:user_item_understanding_continue_old}
    \setlength{\tabcolsep}{1.1mm}{
    \begin{tabular}{llccccccccccc}
    \toprule
    &  & \multicolumn{5}{c}{GLM-4			
} & \multicolumn{2}{c}{Qwen2.5-Ins			
} & \multicolumn{2}{c}{Qwen2.5			
} & \multicolumn{2}{c}{LLaMA-3.2		
} \\
    \cmidrule(lr){3-7}\cmidrule(lr){8-9}\cmidrule(lr){10-11}\cmidrule(lr){12-13}
    Task \& Dataset & Metric & No FT & FT & 25\% & 50\% & 75\%  & No FT & FT & No FT & FT & No FT & FT\\
    \midrule
     \multicolumn{13}{l}{\textbf{{Target User Identification}}} \\
\multirow{5}{*}{Amazon Reviews 2023}
     & BLEU-1 & 0.2039 & 0.4819 & 0.4782 & 0.4856 & 0.4872 & 0.3008 & 0.4751 & 0.2082 & 0.4566 & 0.1254 & 0.4284 \\
     & BLEU-2 & 0.1088 & 0.3314 & 0.3252 & 0.3338 & 0.3341 & 0.1730 & 0.3240 & 0.1149 & 0.3122 & 0.0606 & 0.2892 \\
     & ROUGE-1 & 0.3118 & 0.5152 & 0.5059 & 0.5130 & 0.5151 & 0.3730 & 0.4875 & 0.3157 & 0.4838 & 0.2110 & 0.4782 \\
     & ROUGE-2 & 0.0982 & 0.2552 & 0.2463 & 0.2537 & 0.2549 & 0.1262 & 0.2393 & 0.1036 & 0.2399 & 0.0630 & 0.2330 \\
     & ROUGE-L & 0.1912 & 0.4876 & 0.4770 & 0.4843 & 0.4862 & 0.2009 & 0.3246 & 0.1610 & 0.3139 & 0.0573 & 0.3020 \\
\bottomrule
    \end{tabular}
    }
\end{table*}

\begin{table*}[t]
    \centering
        \caption{Performance on user-item interaction tasks after removing different root tasks using ITDR across four models.}
    \label{tab:user_item_interaction_categories}
    \begin{threeparttable}
    \begin{tabular}{llcccccccc}
    \toprule
    &  & \multicolumn{2}{c}{GLM-4			
} & \multicolumn{2}{c}{Qwen2.5-Ins			
} & \multicolumn{2}{c}{Qwen2.5			
} & \multicolumn{2}{c}{LLaMA-3.2
}\\
    \cmidrule(lr){3-4}\cmidrule(lr){5-6}\cmidrule(lr){7-8}\cmidrule(lr){9-10}
    Task \& Dataset & Metric & w/o UII & w/o UIU& w/o UII & w/o UIU& w/o UII & w/o UIU& w/o UII & w/o UIU\\
    \midrule
\multicolumn{8}{l}{\textbf{{Rating Prediction}}} \\
MovieLens 32M 
 & RMSE & 0.9381 & 1.0856 & 1.1694 & 1.0761 & 1.0228 & 1.0954 & 1.1422 & 1.2661 \\
BookCrossing 
 & RMSE & 1.7941 & 1.7846 & 2.3248 & 1.9491 & 1.9047 & 1.8163 & 1.7800 & 1.7678 \\
Anime Dataset 2023 
 & RMSE & 1.5681 & 1.3856 & 1.7760 & 1.5569 & 1.6997 & 1.4642 & 1.6383 & 1.4339 \\
Yelp
 & RMSE & 1.9365 & 1.9768 & 2.5018 & 2.0077 & 2.0514 & 1.9978 & 2.1714 & 2.0477 \\

\midrule
\multicolumn{8}{l}{\textbf{{Top-K Recommendation}}} \\
\bottomrule
    \end{tabular}
    \begin{tablenotes}
    \item \footnotesize\textsuperscript{*}“UII” means “user-item interaction”, “UIU” means “user-item understanding”.
    \end{tablenotes}
    \end{threeparttable}

\end{table*}

\begin{table*}[t]
    \centering
    \setlength{\tabcolsep}{1mm}{
    \begin{threeparttable}
        \caption{Performance on user-item understanding tasks after removing different root tasks using ITDR across four models.}
    \label{tab:user_item_understanding_categories}
    \begin{tabular}{llcccccccc}
    \toprule
    &  & \multicolumn{2}{c}{GLM-4			
} & \multicolumn{2}{c}{Qwen2.5-Ins			
} & \multicolumn{2}{c}{Qwen2.5			
} & \multicolumn{2}{c}{LLaMA-3.2
} \\
    \cmidrule(lr){3-4}\cmidrule(lr){5-6}\cmidrule(lr){7-8}\cmidrule(lr){9-10}
    Task \& Dataset & Metric & w/o UII & w/o UIU& w/o UII & w/o UIU& w/o UII & w/o UIU& w/o UII & w/o UIU\\
    \midrule
\multicolumn{8}{l}{\textbf{{User Attribute Prediction}}} \\
Last.FM 1K 
 & ACC(country) & 0.5946 & 0.2432 & 0.4595 & 0.2973 & 0.5405 & 0.2162 & 0.3243 & 0.1351 \\
 & ACC(gender) & 0.5135 & 0.1081 & 0.7027 & 0.6486 & 0.3784 & 0.1892 & 0.2162 & 0.1081 \\
 & RMSE(age) & 6.2560 & 6.8260 & 6.0467 & 6.4431 & 7.3695 & 7.6896 & 6.5574 & 6.9862 \\
Last.FM 360K 
 & ACC(country) & 0.4440 & 0.1040 & 0.3880 & 0.2640 & 0.3920 & 0.2600 & 0.4080 & 0.0960 \\
 & ACC(gender) & 0.7440 & 0.2720 & 0.7160 & 0.7240 & 0.7000 & 0.4080 & 0.7200 & 0.4680 \\
\bottomrule
    \end{tabular}
    \begin{tablenotes}
    \item \footnotesize\textsuperscript{*}“UII” means “user-item interaction”, “UIU” means “user-item understanding”.
    \end{tablenotes}
    \end{threeparttable}
    }

\end{table*}

\begin{table*}[t]
    \centering
        \caption{Continued from Table~\ref{tab:user_item_understanding_categories}.}
    \label{tab:user_item_understanding_categories_continue}
    \setlength{\tabcolsep}{1mm}{
    \begin{tabular}{llcccccccc}
    \toprule
    &  & \multicolumn{2}{c}{GLM-4			
} & \multicolumn{2}{c}{Qwen2.5-Ins			
} & \multicolumn{2}{c}{Qwen2.5			
} & \multicolumn{2}{c}{LLaMA-3.2		
} \\
    \cmidrule(lr){3-4}\cmidrule(lr){5-6}\cmidrule(lr){7-8}\cmidrule(lr){9-10}
    Task \& Dataset & Metric & w/o UII & w/o UIU& w/o UII & w/o UIU& w/o UII & w/o UIU& w/o UII & w/o UIU\\
    \midrule
\multicolumn{8}{l}{\textbf{{Target User Identification}}} \\
\multirow{5}{*}{Amazon Reviews 2023}
 & BLEU-1 & 0.4884 & 0.0000 & 0.4707 & 0.0048 & 0.4359 & 0.0481 & 0.3675 & 0.0580 \\
 & BLEU-2 & 0.3362 & 0.0000 & 0.3185 & 0.0023 & 0.2941 & 0.0196 & 0.2467 & 0.0178 \\
 & ROUGE-1 & 0.5161 & 0.0395 & 0.4849 & 0.1146 & 0.4733 & 0.1410 & 0.4757 & 0.1345 \\
 & ROUGE-2 & 0.2555 & 0.0066 & 0.2347 & 0.0283 & 0.2284 & 0.0285 & 0.2293 & 0.0169 \\
 & ROUGE-L & 0.4868 & 0.0373 & 0.3197 & 0.0692 & 0.3013 & 0.0484 & 0.2760 & 0.0905 \\
\bottomrule
    \end{tabular}
    }

\end{table*}

\begin{table*}[t]
    \centering
    \begin{threeparttable}
        \caption{Results for user-item interaction tasks on GLM-4-9B with ITDR removing different subtasks.}
    \label{tab:user_item_interaction_Ablation}
    \begin{tabular}{llcccccccc}
    \toprule
    Task \& Dataset & Metric & \textit{w/o} RP & \textit{w/o} TKR & \textit{w/o} CDR & \textit{w/o} NIR & \textit{w/o} UAP & \textit{w/o} IR & \textit{w/o} TUI & \textit{w/o} TD\\
    \midrule
    \multicolumn{10}{l}{\textbf{{Rating Prediction}}} \\
    MovieLens 32M 
     & RMSE & 1.0271 & 0.9975 & 1.0488 & 0.9022 & 1.0178 & 0.9854 & 0.9381 & 0.9592  \\
    BookCrossing 
     & RMSE & 2.1443 & 1.6402 & 1.7131 & 1.7321 & 1.7020 & 1.7131 & 1.8445 & 1.7907  \\
    Anime Dataset 2023
     & RMSE & 1.5336 & 1.3130 & 1.3900 & 1.3928 & 1.3638 & 1.3711 & 1.3784 & 1.4656 \\
    Yelp
     & RMSE & 2.2282 & 1.9331 & 1.9890 & 1.9702 & 2.0131 & 1.9354 & 2.0402 & 2.0251  \\
    \midrule
    \multicolumn{10}{l}{\textbf{{Top-K Recommendation}}} \\
   Amazon Reviews 2023 
     & N@10 & 0.9085 & 0.6419 & 0.9111 & 0.9037 & 0.9090 & 0.9156 & 0.9086 & 0.9078 \\
\bottomrule
    \end{tabular}
    \begin{tablenotes}
    \item \footnotesize\textsuperscript{*}“RP” stands for “Rating Prediction”, “TKR” means “Top-K Recommendation”, “CDR” means “Cross-Domain Recommendation”, “NIR” refers to “Next Item Recommendation”, “UAP” indicates “User Attribute Prediction”, “IR” indicates “Interest Recognition”, “TUI” denotes “Target User Identification”, and "TD" represents "Task Description".
    \end{tablenotes}
    \end{threeparttable}
\end{table*}

\begin{table*}[t]
    \centering
    \setlength{\tabcolsep}{1mm}{
    \begin{threeparttable}
        \caption{Results for user-item understanding tasks on GLM-4-9B with ITDR removing different subtasks.}
    \label{tab:user_item_understanding_Ablation1}
    \begin{tabular}{llcccccccc}
    \toprule
    Task \& Dataset & Metric & \textit{w/o} RP & \textit{w/o} TKR & \textit{w/o} CDR & \textit{w/o} NIR & \textit{w/o} UAP & \textit{w/o} IR & \textit{w/o} TUI & \textit{w/o} TD\\
    \midrule
    \multicolumn{10}{l}{\textbf{{User Attribute Prediction}}} \\
    Last.FM 1K 
     & ACC(country) & 0.5405 & 0.6486 & 0.5676 & 0.6216 & 0.2432 & 0.5676 & 0.4865 & 0.5946 \\
     & ACC(gender) & 0.4595 & 0.6216 & 0.6216 & 0.4865 & 0.3243 & 0.5946 & 0.5135 & 0.5135 \\
     & RMSE(age) & 6.4512 & 6.5533 & 6.8569 & 6.1045 & 8.2658 & 6.0589 & 6.1298 & 6.8821 \\
    Last.FM 360K 
     & ACC(country) & 0.4600 & 0.4720 & 0.4480 & 0.4320 & 0.2240 & 0.4360 & 0.4680 & 0.4800 \\
     & ACC(gender) & 0.7520 & 0.7240 & 0.7520 & 0.7400 & 0.6600 & 0.7400 & 0.7320 & 0.7160 \\
     & RMSE(age) & 8.9140 & 9.0841 & 8.6457 & 8.9904 & 10.1927 & 8.4669 & 8.8367 & 8.7964 \\
    Anime Dataset 2023
     & ACC(location) & 0.0320 & 0.0480 & 0.0600 & 0.0360 & 0.0080 & 0.0440 & 0.0400 & 0.0400 \\
     & ACC(gender) & 0.8120 & 0.8080 & 0.8240 & 0.7960 & 0.6560 & 0.8160 & 0.8440 & 0.8320 \\
     & RMSE(Birth year) & 4.6152 & 4.6673 & 4.3474 & 4.5760 & 6.8148 & 4.6411 & 4.4231 & 4.3433 \\
    BookCrossing
     & ACC(location) & 0.0560 & 0.0720 & 0.0600 & 0.0400 & 0.0000 & 0.0520 & 0.0680 & 0.0680 \\
     & RMSE(age) & 16.5817 & 15.6690 & 16.4448 & 15.5870 & 18.2569 & 15.6729 & 15.5358 & 15.1733 \\
    MovieLens 1M
     & ACC(age group) & 0.3750 & 0.3958 & 0.3583 & 0.3875 & 0.3667 & 0.3750 & 0.4000 & 0.3708 \\
     & ACC(gender) & 0.7542 & 0.7792 & 0.7417 & 0.7750 & 0.5375 & 0.7250 & 0.7458 & 0.7375 \\
     & ACC(occupation) & 0.1292 & 0.1458 & 0.1500 & 0.1292 & 0.0000 & 0.1542 & 0.1168 & 0.0958 \\
    \midrule
    \multicolumn{10}{l}{\textbf{{Interest Recognition}}} \\
    Amazon Books Reviews
     & BLEU-1 & 0.3571 & 0.3502 & 0.3446 & 0.3515 & 0.3481 & 0.1782 & 0.3549 & 0.3604 \\
     & BLEU-2 & 0.2235 & 0.2174 & 0.2156 & 0.2174 & 0.2164 & 0.0865 & 0.2177 & 0.2255 \\
     & ROUGE-1 & 0.4185 & 0.4138 & 0.4119 & 0.4156 & 0.4135 & 0.2782 & 0.4135 & 0.4210 \\
     & ROUGE-2 & 0.1811 & 0.1754 & 0.1753 & 0.1753 & 0.1754 & 0.0758 & 0.1731 & 0.1799 \\
     & ROUGE-L & 0.3941 & 0.3890 & 0.3865 & 0.3899 & 0.3890 & 0.2598 & 0.3897 & 0.3944 \\
    Anime Dataset 2023
     & BLEU-1 & 0.4424 & 0.4449 & 0.4455 & 0.4486 & 0.4386 & 0.2371 & 0.4394 & 0.4379 \\
     & BLEU-2 & 0.3337 & 0.3362 & 0.3371 & 0.3336 & 0.3290 & 0.1230 & 0.3316 & 0.3308 \\
     & ROUGE-1 & 0.5435 & 0.5440 & 0.5400 & 0.5455 & 0.5350 & 0.3120 & 0.5388 & 0.5409 \\
     & ROUGE-2 & 0.3040 & 0.3056 & 0.3053 & 0.2982 & 0.2965 & 0.0857 & 0.3003 & 0.3025 \\
     & ROUGE-L & 0.5084 & 0.5080 & 0.5081 & 0.5101 & 0.4996 & 0.2816 & 0.5046 & 0.5057 \\
    Steam
     & BLEU-1 & 0.4676 & 0.4626 & 0.4626 & 0.4670 & 0.4730 & 0.2126 & 0.4582 & 0.4631 \\
     & BLEU-2 & 0.3166 & 0.3131 & 0.3128 & 0.3145 & 0.3221 & 0.1208 & 0.3091 & 0.3123 \\
     & ROUGE-1 & 0.5000 & 0.4945 & 0.4900 & 0.4955 & 0.4993 & 0.3482 & 0.4928 & 0.4964 \\
     & ROUGE-2 & 0.2395 & 0.2365 & 0.2345 & 0.2347 & 0.2422 & 0.1117 & 0.2332 & 0.2359 \\
     & ROUGE-L & 0.4669 & 0.4618 & 0.4586 & 0.4614 & 0.4666 & 0.3183 & 0.4590 & 0.4623 \\
    \bottomrule
    \end{tabular}
    \begin{tablenotes}
    \item \footnotesize\textsuperscript{*}“RP” stands for “Rating Prediction”, “TKR” means “Top-K Recommendation”, “CDR” means “Cross-Domain Recommendation”, “NIR” refers to “Next Item Recommendation”, “UAP” indicates “User Attribute Prediction”, “IR” indicates “Interest Recognition”, “TUI” denotes “Target User Identification”, and "TD" represents "Task Description".
    \end{tablenotes}
    \end{threeparttable}
    }
\end{table*}

\begin{table*}[t]
    \centering
        \caption{Continued from Table~\ref{tab:user_item_understanding_Ablation1}.}
    \label{tab:user_item_understanding_Ablation2}
    \begin{tabular}{llcccccccc}
    \toprule
    Task \& Dataset & Metric & \textit{w/o} RP & \textit{w/o} TKR & \textit{w/o} CDR & \textit{w/o} NIR & \textit{w/o} UAP & \textit{w/o} IR & \textit{w/o} TUI & \textit{w/o} TD\\
    \midrule
    \multicolumn{10}{l}{\textbf{{Target User Identification}}} \\
\multirow{5}{*}{Amazon Reviews 2023} 
     & BLEU-1 & 0.4892 & 0.4858 & 0.4920 & 0.4856 & 0.4909 & 0.4827 & 0.0363 & 0.4866 \\
     & BLEU-2 & 0.3381 & 0.3328 & 0.3387 & 0.3329 & 0.3382 & 0.3320 & 0.0184 & 0.3353 \\
     & ROUGE-1 & 0.5203 & 0.5138 & 0.5182 & 0.5116 & 0.5165 & 0.5137 & 0.2261 & 0.5172 \\
     & ROUGE-2 & 0.2603 & 0.2536 & 0.2588 & 0.2530 & 0.2575 & 0.2552 & 0.0547 & 0.2580 \\
     & ROUGE-L & 0.4899 & 0.4856 & 0.4887 & 0.4834 & 0.4875 & 0.4859 & 0.1995 & 0.4870 \\
     \midrule
\multirow{5}{*}{BookCrossing} 
     & BLEU-1 & 0.5088 & 0.5125 & 0.5052 & 0.5090 & 0.5132 & 0.5096 & 0.0073 & 0.5072 \\
     & BLEU-2 & 0.3605 & 0.3601 & 0.3552 & 0.3576 & 0.3639 & 0.3580 & 0.0039 & 0.3582 \\
     & ROUGE-1 & 0.5377 & 0.5429 & 0.5363 & 0.5386 & 0.5454 & 0.5412 & 0.1663 & 0.5384 \\
     & ROUGE-2 & 0.2783 & 0.2763 & 0.2734 & 0.2744 & 0.2830 & 0.2747 & 0.0419 & 0.2771 \\
     & ROUGE-L & 0.5063 & 0.5100 & 0.5053 & 0.5062 & 0.5143 & 0.5091 & 0.1467 & 0.5070 \\
     \midrule
\multirow{5}{*}{Anime Dataset 2023}
     & BLEU-1 & 0.5463 & 0.5469 & 0.5497 & 0.5461 & 0.5520 & 0.5452 & 0.0064 & 0.5492 \\
     & BLEU-2 & 0.4074 & 0.4067 & 0.4065 & 0.4085 & 0.4110 & 0.4040 & 0.0037 & 0.4098 \\
     & ROUGE-1 & 0.5809 & 0.5831 & 0.5845 & 0.5841 & 0.5885 & 0.5825 & 0.1515 & 0.5859 \\
     & ROUGE-2 & 0.3275 & 0.3270 & 0.3251 & 0.3324 & 0.3306 & 0.3244 & 0.0344 & 0.3309 \\
     & ROUGE-L & 0.5502 & 0.5497 & 0.5516 & 0.5534 & 0.5572 & 0.5508 & 0.1315 & 0.5544 \\
     \midrule
\multirow{5}{*}{Steam}
     & BLEU-1 & 0.5104 & 0.5144 & 0.5090 & 0.5144 & 0.5158 & 0.5125 & 0.0942 & 0.5121 \\
     & BLEU-2 & 0.3686 & 0.3711 & 0.3641 & 0.3699 & 0.3706 & 0.3703 & 0.0509 & 0.3697 \\
     & ROUGE-1 & 0.5389 & 0.5401 & 0.5355 & 0.5409 & 0.5367 & 0.5389 & 0.2977 & 0.5396 \\
     & ROUGE-2 & 0.2866 & 0.2868 & 0.2787 & 0.2852 & 0.2833 & 0.2866 & 0.0857 & 0.2865 \\
     & ROUGE-L & 0.5087 & 0.5101 & 0.5044 & 0.5086 & 0.5067 & 0.5095 & 0.2597 & 0.5097 \\
     \midrule
\multirow{5}{*}{Yelp}
     & BLEU-1 & 0.4824 & 0.4819 & 0.4872 & 0.4826 & 0.4842 & 0.4848 & 0.0786 & 0.4849 \\
     & BLEU-2 & 0.3292 & 0.3312 & 0.3325 & 0.3293 & 0.3327 & 0.3310 & 0.0404 & 0.3318 \\
     & ROUGE-1 & 0.5149 & 0.5185 & 0.5182 &  0.5188 & 0.5185 & 0.5184 & 0.2552 & 0.5192 \\
     & ROUGE-2 & 0.2523 & 0.2575 & 0.2534 & 0.2541 & 0.2572 & 0.2540 & 0.0676 &  0.2563 \\
     & ROUGE-L & 0.4870 & 0.4913 & 0.4899 & 0.4896 & 0.4903 & 0.4913 & 0.2299 & 0.4901 \\
     \midrule
\multirow{5}{*}{MicroLens}
     & BLEU-1 & 0.4996 & 0.5023 & 0.4976 & 0.5047 & 0.5011 & 0.5001 & 0.0025 & 0.5049 \\
     & BLEU-2 & 0.3581 & 0.3617 & 0.3560 & 0.3640 & 0.3614 & 0.3600 & 0.0010 & 0.3628 \\
     & ROUGE-1 & 0.5331 & 0.5360 & 0.5317 & 0.5399 & 0.5396 & 0.5369 & 0.1295 & 0.5393 \\
     & ROUGE-2 & 0.2809 & 0.2850 & 0.2797 & 0.2871 & 0.2872 & 0.2843 & 0.0272 & 0.2848 \\
     & ROUGE-L & 0.5036 & 0.5082 & 0.5026 & 0.5103 & 0.5109 & 0.5094 & 0.1147 & 0.5086 \\
     \midrule
\multirow{5}{*}{PixelRec}
     & BLEU-1 & 0.4843 & 0.4871 & 0.4869 & 0.4874 & 0.4866 & 0.4849 & 0.0012 & 0.4896 \\
     & BLEU-2 & 0.3436 & 0.3445 & 0.3452 & 0.3453 & 0.3437 & 0.3438 & 0.0006 & 0.3486 \\
     & ROUGE-1 & 0.5203 & 0.5237 & 0.5221 & 0.5242 & 0.5226 & 0.5222 & 0.1243 & 0.5271 \\
     & ROUGE-2 & 0.2656 & 0.2663 & 0.2676 & 0.2676 & 0.2649 & 0.2665 & 0.0281 & 0.2707 \\
     & ROUGE-L & 0.4885 & 0.4926 & 0.4920 & 0.4921 & 0.4904 & 0.4912 & 0.1093 & 0.4955 \\
     \midrule
\multirow{5}{*}{MovieLens 32M}
     & BLEU-1 & 0.5311 & 0.5232 & 0.5268 & 0.5210 & 0.5227 & 0.5196 & 0.0021 & 0.5247 \\
     & BLEU-2 & 0.3876 & 0.3806 & 0.3830 & 0.3758 & 0.3796 &  0.3763 & 0.0013 & 0.3806 \\
     & ROUGE-1 & 0.5529 & 0.5482 & 0.5504 & 0.5456 & 0.5493 & 0.5454 & 0.1375 & 0.5502 \\
     & ROUGE-2 & 0.2997 & 0.2950 & 0.2956 & 0.2890 & 0.2945 & 0.2915 & 0.0331 & 0.2940 \\
     & ROUGE-L & 0.5154 & 0.5102 & 0.5116 & 0.5054 & 0.5109 & 0.5081 & 0.1186 & 0.5111 \\
     \midrule
\multirow{5}{*}{MIND} 
     & BLEU-1 & 0.5578 & 0.5571 & 0.5574 & 0.5587 & 0.5520 & 0.5534 & 0.0010 & 0.5627 \\
     & BLEU-2 & 0.4034 & 0.4029 & 0.3993 & 0.4043 & 0.3972 & 0.4004 & 0.0006 & 0.4072 \\
     & ROUGE-1 & 0.6104 & 0.6098 & 0.6097 & 0.6096 & 0.6084 & 0.6108 & 0.1339 & 0.6153 \\
     & ROUGE-2 & 0.3194 & 0.3188 & 0.3131 & 0.3198 & 0.3151 & 0.3199 & 0.0351 & 0.3233 \\
     & ROUGE-L & 0.5777 & 0.5773 & 0.5752 & 0.5780 & 0.5754 & 0.5789 & 0.1215 & 0.5824 \\
    \bottomrule
    \end{tabular}
\end{table*}

\begin{table*}[t]
    \centering
        \caption{Experimental results of closed source LLMs on user-item interaction tasks.}
    \label{tab:user_item_interaction_gpt_gemini}
    \begin{tabular}{llcc}
    \toprule
    Task \& Dataset & Metric & GPT-3.5-Turbo &  Gemini 2.5 Flash \\
    \midrule
    \multicolumn{2}{l}{\textbf{{Rating Prediction}}} \\
    MovieLens 32M 
     & RMSE & 0.9775 & 0.9644 \\
BookCrossing 
     & RMSE & 1.6583 & 1.8947 \\
Anime Dataset 2023
     & RMSE & 1.4000 & 1.4156 \\
Yelp
     & RMSE & 1.5602 & 2.1422 \\
\midrule
\multicolumn{2}{l}{\textbf{{Top-K Recommendation}}} \\
Amazon Reviews 2023 
     & N@10 & 0.8630 & 0.8682 \\
     & H@1 & 0.7797 & 0.8093 \\
     & P@5 & 0.6576 & 0.6941 \\
MovieLens 32M
     & N@10 & 0.9696 & 0.9685 \\
     & H@1 & 0.9800 & 0.9440 \\
     & P@5 & 0.8920 & 0.8856 \\
MicroLens
     & N@10 & 0.7854 & 0.8009 \\
     & H@1 & 0.6760 & 0.7200 \\
     & P@5 & 0.5528 & 0.6000 \\
PixelRec
     & N@10 & 0.7241 & 0.7687 \\
     & H@1 & 0.5840 & 0.7000 \\
     & P@5 & 0.4752 & 0.5456 \\
Amazon Books Reviews
     & N@10 & 0.9661 & 0.9239 \\
     & H@1 & 0.9600 & 0.9320 \\
     & P@5 & 0.8080 & 0.8376 \\
MIND 
     & N@10 & 0.7598 & 0.7523 \\
     & H@1 & 0.5960 & 0.6440 \\
     & P@5 & 0.5237 & 0.5304 \\
Anime Dataset 2023
     & N@10 & 0.9790 & 0.9527 \\
     & H@1 & 0.9920 & 0.9200 \\
     & P@5 & 0.8808 & 0.8616 \\
\midrule
\multicolumn{2}{l}{\textbf{Cross-Domain Recommendation}} \\
Amazon Reviews 2023 
     & N@10 & 0.8111 & 0.8335 \\
     & H@1 & 0.6048 & 0.7605 \\
     & P@5 & 0.5928 & 0.6419 \\
\midrule
\multicolumn{2}{l}{\textbf{Next Item Recommendation}} \\
Amazon Reviews 2023
     & H@1 & 0.3723 & 0.5801 \\
MicroLens
     & H@1 & 0.3440 & 0.5280 \\
MovieLens 32M 
     & H@1 & 0.3960 & 0.4360 \\
PixelRec 
     & H@1 & 0.2880 & 0.2400 \\
    \bottomrule
    \end{tabular}
\end{table*}

\begin{table*}[t]
    \centering
        \caption{Experimental results of closed source LLMs on user-item understanding tasks.}
    \label{tab:user_item_understanding_gpt_gemini}
    \begin{tabular}{llcc}
    \toprule
    Task \& Dataset & Metric & GPT-3.5-Turbo &  Gemini 2.5 Flash \\
    \midrule
\multicolumn{2}{l}{\textbf{{User Attribute Prediction}}} \\
Last.FM 1K 
     & ACC(country) & 0.1622 & 0.3514 \\
     & ACC(gender) & 0.5405 & 0.6757 \\
     & RMSE(age) & 5.0929 & 7.7127 \\
Last.FM 360K 
     & ACC(country) & 0.2440 & 0.4080 \\
     & ACC(gender) & 0.5360 & 0.5280 \\
     & RMSE(age) & 9.1007 & 11.2361 \\
Anime Dataset 2023
     & ACC(location) & 0.0240 & 0.0000 \\
     & ACC(gender) & 0.5560 & 0.6840 \\
     & RMSE(Birth year) & 5.4794 & 5.5656 \\
BookCrossing
     & ACC(location) & 0.0800 & 0.0800 \\
     & RMSE(age) & 14.5382 & 15.6048 \\
MovieLens 1M
     & ACC(age group) & 0.3167 & 0.0667 \\
     & ACC(gender) & 0.7208 & 0.7250 \\
     & ACC(occupation) & 0.0083 & 0.0167 \\
\midrule
\multicolumn{2}{l}{\textbf{{Interest Recognition}}} \\
Amazon Books Reviews
     & BLEU-1 & 0.1551 & 0.1030 \\
     & BLEU-2 & 0.0675 & 0.0386 \\
     & ROUGE-1 & 0.2098 & 0.1706 \\
     & ROUGE-2 & 0.0420 & 0.0245 \\
     & ROUGE-L & 0.1368 & 0.0975 \\
Anime Dataset 2023
     & BLEU-1 & 0.1452 & 0.1412 \\
     & BLEU-2 & 0.0830 & 0.0559 \\
     & ROUGE-1 & 0.2560 & 0.2032 \\
     & ROUGE-2 & 0.0771 & 0.0332 \\
     & ROUGE-L & 0.1662 & 0.1497 \\
Steam
     & BLEU-1 & 0.1753 & 0.1014 \\
     & BLEU-2 & 0.1046 & 0.0469 \\
     & ROUGE-1 & 0.3075 & 0.2297 \\
     & ROUGE-2 & 0.1177 & 0.0523 \\
     & ROUGE-L & 0.2032 & 0.1335 \\
    \bottomrule
    \end{tabular}
\end{table*}

\begin{table*}[t]
    \centering
        \caption{Continued from Table~\ref{tab:user_item_understanding_gpt_gemini}.}
    \label{tab:user_item_understanding_gpt_gemini_continue}
    \begin{tabular}{llcc}
    \toprule
    Task \& Dataset & Metric & GPT-3.5-Turbo &  Gemini 2.5 Flash \\
    \midrule
\multicolumn{2}{l}{\textbf{{Target User Identification}}} \\
\multirow{5}{*}{Amazon Reviews 2023} 
     & BLEU-1 & 0.3182 & 0.2862 \\
     & BLEU-2 & 0.1811 & 0.1474 \\
     & ROUGE-1 & 0.3644 & 0.3277 \\
     & ROUGE-2 & 0.1275 & 0.0946 \\
     & ROUGE-L & 0.2229 & 0.1854 \\
     \midrule
\multirow{5}{*}{BookCrossing} 
     & BLEU-1 & 0.2481 & 0.2751 \\
     & BLEU-2 & 0.1512 & 0.1455 \\
     & ROUGE-1 & 0.3786 & 0.3436 \\
     & ROUGE-2 & 0.1433 & 0.1013 \\
     & ROUGE-L & 0.2323 & 0.2008 \\
     \midrule
\multirow{5}{*}{Anime Dataset 2023}
     & BLEU-1 & 0.2983 & 0.2499 \\
     & BLEU-2 & 0.1774 & 0.1303 \\
     & ROUGE-1 & 0.3799 & 0.3267 \\
     & ROUGE-2 & 0.1382 & 0.0922 \\
     & ROUGE-L & 0.2311 & 0.1886 \\
     \midrule
\multirow{5}{*}{Steam}
     & BLEU-1 & 0.3050 & 0.2267 \\
     & BLEU-2 & 0.1783 & 0.1130 \\
     & ROUGE-1 & 0.3703 & 0.3027 \\
     & ROUGE-2 & 0.1322 & 0.0779 \\
     & ROUGE-L & 0.2241 & 0.1713 \\
     \midrule
\multirow{5}{*}{Yelp}
     & BLEU-1 & 0.3122 & 0.2694 \\
     & BLEU-2 & 0.1784 & 0.1419 \\
     & ROUGE-1 & 0.3622 & 0.3278 \\
     & ROUGE-2 & 0.1238 & 0.0960 \\
     & ROUGE-L & 0.2113 & 0.1914 \\
     \midrule
\multirow{5}{*}{MicroLens}
     & BLEU-1 & 0.1700 & 0.3419 \\
     & BLEU-2 & 0.1185 & 0.1996 \\
     & ROUGE-1 & 0.3758 & 0.3722 \\
     & ROUGE-2 & 0.1806 & 0.1336 \\
     & ROUGE-L & 0.2569 & 0.2299 \\
     \midrule
\multirow{5}{*}{PixelRec}
     & BLEU-1 & 0.1893 & 0.3338 \\
     & BLEU-2 & 0.1159 & 0.1938 \\
     & ROUGE-1 & 0.3416 & 0.3675 \\
     & ROUGE-2 & 0.1303 & 0.1304 \\
     & ROUGE-L & 0.2093 & 0.2248 \\
     \midrule
\multirow{5}{*}{MovieLens 32M}
     & BLEU-1 & 0.2281 & 0.2626 \\
     & BLEU-2 & 0.1358 & 0.1456 \\
     & ROUGE-1 & 0.3478 & 0.3382 \\
     & ROUGE-2 & 0.1288 & 0.1098 \\
     & ROUGE-L & 0.2079 & 0.2096 \\
     \midrule
\multirow{5}{*}{MIND}
     & BLEU-1 & 0.2839 & 0.3327 \\
     & BLEU-2 & 0.1705 & 0.1875 \\
     & ROUGE-1 & 0.4134 & 0.4093 \\
     & ROUGE-2 & 0.1455 & 0.1297 \\
     & ROUGE-L & 0.2378 & 0.2436 \\
    \bottomrule
    \end{tabular}
\end{table*}

\begin{table*}[t]
    \centering
        \caption{A data example of the rating prediction task. Data source: Anime Dataset 2023.}
    \label{tab:RP_EXP}
    \small
    \begin{tabular}{p{.95\linewidth}}
    \toprule
        \textbf{Rating Prediction}  \\
    \midrule
        \textbf{Task Description}:
        \\
        The objective of the rating prediction task is to thoroughly explore users' preferences toward items (e.g., movies, products, music, books, etc.) based on their historical rating records. By extracting insights from the rating data to identify the types of items users prefer or dislike, the task aims to accurately predict users' ratings for new items.
        \\
        \textbf{Input}:
        \\
        A user has watched and rated the following anime with ratings: Bleach (7), Naruto: Shippuuden (6), Bishoujo Senshi Sailor Moon (7), Bishoujo Senshi Sailor Moon S: Kaguya-hime no Koibito (4), Bishoujo Senshi Sailor Moon SuperS (4), Bishoujo Senshi Sailor Moon SuperS Specials (4), Bleach Movie 1: Memories of Nobody (7), Bleach Movie 3: Fade to Black - Kimi no Na wo Yobu (6), Bleach: Memories in the Rain (6), Bleach: The Sealed Sword Frenzy (6). Predict his rating for Chrno Crusade.
        \\
        \midrule
        \textbf{Output}:
        \\
        9
        \\
    \bottomrule
    \end{tabular}
\end{table*}

\begin{table*}[t]
    \centering
        \caption{A data example of the top k recommendation task. Data source: PixelRec.}
    \label{tab:TKR_EXP}
    \small
    \begin{tabular}{p{.95\linewidth}}
    \toprule
        \textbf{Top-K Recommendation}  \\
    \midrule
        \textbf{Task Description}:
        \\
        The primary goal of Top-K recommendation is to select the K items most likely to attract the user from a candidate set based on their historical behavior or preferences, and present them in a ranked list.
        \\
        \textbf{Input}:
        \\
      Given a user's short video watching records in chronological order as follows:\\
      1. How to get the most out of your favorite vegetables in the world? Why are all American vegetables the same?\\
      2. Qingming Festival in Xi'an storm storm eating, get up first drink eight bowls of wrestling bowl wine!\\
      3. The first time I made bean flowers in the countryside, Wang Gang asked me to challenge the spiciest seven-star pepper!\\
      4. Do American college students look for their families for living expenses and tuition? Street interviews with Americans!\\
      5. [European and American group portrait] "You are the hopeless lover I hide under the great friendship."\\
      Based on the watching history, please rank the short video in the candidate list by the likelihood of the user watching them next from highest to lowest.\\
      Candidate list:\\
      1. Game of Thrones Male Character Homo Popularity Rankings\\
      2. The fat woman at home, forcing her pregnant daughter to divorce to take care of herself, stealing food to lose weight and scolding the doctor!\\
      3. Free to receive the original price of Mars Survival Deluxe Edition, come to white whoring hi plus one!\\
      4. Chemistry "Stalactites"\\
      5. [Emma Slippery Won] Facing the straight ball sweet girl, he is really panicked!\\
      6. All up, you must know this niche and good treasure app!\\
      7. Zeta Ultraman meets all generations of Ultraman\\
      8. Game of Thrones: How many times did Haima Raijin laugh at the most hepatic collection on the B station?\\
      9. I, Killsumaru, have no one to protect.\\
      10. President Seahorse
        \\
        \midrule
        \textbf{Output}:
        \\
        1. Game of Thrones: How many times did Haima Raijin laugh at the most hepatic collection on the B station?\\
        2. Game of Thrones Male Character Homo Popularity Rankings\\
        3. President Seahorse\\
        4. I, Killsumaru, have no one to protect.\\
        5. Zeta Ultraman meets all generations of Ultraman\\
        6. The fat woman at home, forcing her pregnant daughter to divorce to take care of herself, stealing food to lose weight and scolding the doctor!\\
        7. [Emma Slippery Won] Facing the straight ball sweet girl, he is really panicked!\\
        8. Free to receive the original price of Mars Survival Deluxe Edition, come to white whoring hi plus one!\\
        9. All up, you must know this niche and good treasure app!\\
        10. Chemistry "Stalactites"\\
    \bottomrule
    \end{tabular}
\end{table*}

\begin{table*}[t]
    \centering
        \caption{A data example of the cross-domain recommendation task. Data source: Amazon Review 2023.}
    \label{tab:CDR_EXP}
    \small
    \begin{tabular}{p{.95\linewidth}}
    \toprule
        \textbf{Cross-Domain Recommendation}  \\
    \midrule
        \textbf{Task Description}:
        \\
        The goal of cross-domain recommendation is to predict a user's behavior in the target domain based on their historical behavior in the source domain, thereby achieving effective utilization of cross-domain information and improving recommendation performance.
        \\
        
        \textbf{Input}:
        \\
           A user's purchase history in the beauty domain is as follows:\\
           1. FIXBODY Boar Bristle Hair Brush (Wooden)\\
           2. Hair Scrunchies for Women \& Girls , 25PCS Hair Scrunchies Accessories Included Velvet, Chiffon, Flower Printed, Silk Hair Ties Band \& Hair Bun Maker, Elastic Hair Scrunchies Perfect for Hair Bun, Ponytail Holder\\
           3. Liquid Foundation Cream, Concealer Cover Cream, BB Cream, Flawless Colour Changing Foundation Makeup Base Nude Face Liquid Cover Concealer, Universal for ALL Skin Types, SPF 15 Shell, 1.41 Fl Oz\\
           4. Liquid Foundation,Foundation Cream,Flawless Colour Changing Foundation,Hides Wrinkles \& Lines,BB Cream,Covering Imperfections Liquid Complete Foundation Cover,Universal Shade for ALL Skin Makeup\\
           5. Morovan 8 Colors Gel Nail Polish Kit with 48W UV LED Light Nail Lamp Pink White Blue Purple Colors For Women With Base and Top Coat Full DIY Professional Gel Manicure Nail Tools for Home and Salon Use,\\
           and in the fashion domain is as follows:
           1. MITALOO Push Up Strapless Sticky Adhesive Invisible Backless Bras Plunge Reusable Magic Bra for Women Beige\\
           2. ENIDMIL Womens Soft Cotton Pajama Set Short Sleeve Top and Long Pants Bottom Sleepwear Loungewear PJ Set Pajamas for Women (Viola Purple, Large)\\
           3. Fettero Women Simple 14K Gold Plated Chain Delicate Single Freshwater Pearls Double Layer Choker Necklace\\
           4. Rave Hood 3D Bandana Gaiter Headwear Headband Magic Scarf Seamless 2019 Improved for Fishing, Man Women Running, Hiking, Motorcycle\\
           5. Three Dots Pendants Necklaces,14k Gold Plated 3 Dot Circle Disks Chain Tiny Triple Round Flat Bead Dainty Handmade Pendants Necklaces Jewelry Gifts for Women Girls.\\
           Based on this information, please rank the following list of products in the fashion domain in order of the likelihood that the user will purchase them next, from highest to lowest. Product list:\\
           1. BSJELL Straw Earrings Woven Handmade Rattan Geometric Drop Earrings Hammered Disc Stud Wicker Bohemian Lightweight Earrings for Women (Teardop)\\
           2. BIADANI Women Versatile Long Sleeve Open Front Drape Front Pockets Cardigan Hacci Pink XX-Large\\
           3. Stainless Steel Men Cuff Link Cuff Links Shirt Set Tile\\
           4. Tassel Fringe Earrings Drop Dangle Earring for Women Girls Black by THSjewelry\\
           5. TOBWIZU Sports Bras for Women,Seamless Sleep Bra with Removable Pads Yoga Gym Activity Everyday Wear\\
           6. DEDU MusBema Kids Casual Lightweight Breathable Sneakers Easy Walk Sport Shoes for Boys Girls White\\
           7. Wannpoty Women's Summer Dress Sleeveless Casual Loose Button T-Shirt with Double Pockets (Large, Dark Green)\\
           8. Oriental Pearl Mama Shirts Women Girl Mom Letter Printed T-Shirt Mama Graphic Tee Casual Short Sleeve Tops Tee (Pink2, XXL)\\
           9. GemsChest 925 Sterling Silver Ring for Women Round Cubic Zirconia 3-Stone Engagement Wedding Jewelry Hola CZ Bridal Ring\\
           10. black Crossbody bags for women Lightweight Medium Classic Bag with tassel and Zipper Pocket Adjustable Strap PU leather.
        \\
        \midrule
        \textbf{Output}:
        \\
        1. BSJELL Straw Earrings Woven Handmade Rattan Geometric Drop Earrings Hammered Disc Stud Wicker Bohemian Lightweight Earrings for Women (Teardop)\\
        2. Tassel Fringe Earrings Drop Dangle Earring for Women Girls Black by THSjewelry\\
        3. black Crossbody bags for women Lightweight Medium Classic Bag with tassel and Zipper Pocket Adjustable Strap PU leather\\
        4. Wannpoty Women's Summer Dress Sleeveless Casual Loose Button T-Shirt with Double Pockets (Large, Dark Green)\\
        5. GemsChest 925 Sterling Silver Ring for Women Round Cubic Zirconia 3-Stone Engagement Wedding Jewelry Hola CZ Bridal Ring\\
        6. TOBWIZU Sports Bras for Women,Seamless Sleep Bra with Removable Pads Yoga Gym Activity Everyday Wear\\
        7. DEDU MusBema Kids Casual Lightweight Breathable Sneakers Easy Walk Sport Shoes for Boys Girls White\\
        8. BIADANI Women Versatile Long Sleeve Open Front Drape Front Pockets Cardigan Hacci Pink XX-Large\\
        9. Oriental Pearl Mama Shirts Women Girl Mom Letter Printed T-Shirt Mama Graphic Tee Casual Short Sleeve Tops Tee (Pink2, XXL)\\
        10. Stainless Steel Men Cuff Link Cuff Links Shirt Set Tile \\
    \bottomrule
    \end{tabular}

\end{table*}

\begin{table*}[ht]
    \centering
        \caption{A data example of the next item recommendation task. Data source: MicroLens.}
    \label{tab:NIR_EXP}
    \small
    \begin{tabular}{p{.95\linewidth}}
    \toprule
        \textbf{Next Item Recommendation}  \\
    \midrule
        \textbf{Task Description}:
        \\
        The primary goal of next-item recommendation is to predict the item that a user is most likely to interact with or be interested in next, based on their historical behavior sequence (such as clicks, purchases, or views).
        \\
        \textbf{Input}:
        \\
        Given that a short video user has watched the following videos in chronological order: ['1.planted sweet potato seedlings after the rain:Delicacy', 
        '2.the field of cowpeas fast old:Delicacy', 
        '3.under the blue sky and white clouds small Viagra with Xiang sister to touch the crunch:Delicacy', 
        '4.far married daughter-in-law pregnant want to eat chicken:Daily Sharing', 
        '5.3 years old cute baby want to super restaurant:Daily Sharing'], 
        please select the next video the user is most likely to watch from the candidate list: 
        ['(follow-up) Yesterday:Daily Sharing', 
        'Let's go! \# Bear-infested \# quadratic nova:Anime', 
        'the sixth set of Diorman princess:Film and Television Works', 
        'in order to make snacks for her son-in-law:Daily Sharing', 
        'Hongli for the rescue of the green Ying does not put the empress dowager access to Cining Palace:Film and Television Works'] 
        \\
        \midrule
        \textbf{Output}:
        \\
        in order to make snacks for her son-in-law \\
    \bottomrule
    \end{tabular}

\end{table*}

\begin{table*}[t]
    \centering
        \caption{A data example of the user attribute prediction task. Data source: MovieLens 1M.}
    \label{tab:UAP_EXP}
    \small
    \begin{tabular}{p{.95\linewidth}}
    \toprule
        \textbf{User Attribute Prediction}  \\
    \midrule
        \textbf{Task Description}:
        \\
        The objective of attribute prediction is to infer implicitly expressed static attributes (e.g., age, gender, income, education level) from user behavioral data (e.g., browsing, purchases, social interactions), thereby completing fundamental dimensions of user profiles to support differentiated recommendation strategies.
        \\
        \textbf{Input}:
        \\
        The user's movie preferences are as follows:
        \\
        He/She likes watching Dead Man Walking (1995): Drama, Dead Poets Society (1989): Drama, Drugstore Cowboy (1989): Crime\textbar Drama, The Graduate (1967): Drama\textbar Romance, North by Northwest (1959): Drama\textbar Thriller, The Insider (1999): Drama, 12 Angry Men (1957): Drama, A Christmas Story (1983): Comedy\textbar Drama, On the Waterfront (1954): Crime\textbar Drama, The Shawshank Redemption (1994): Drama, The Silence of the Lambs (1991): Drama\textbar Thriller, To Kill a Mockingbird (1962): Drama, The Boat(Das Boot) (1981): Action\textbar Drama\textbar War, 2001: A Space Odyssey (1968): Drama\textbar Mystery\textbar Sci-Fi\textbar Thriller, Midnight Cowboy (1969): Drama, Serpico (1973): Crime\textbar Drama, The Wizard of Oz (1939): Adventure\textbar Children's\textbar Drama\textbar Musical, American Beauty (1999): Comedy\textbar Drama, Citizen Kane (1941): Drama, Casablanca (1942): Drama\textbar Romance\textbar War.
        \\
        Based on the user's movie preferences, predict his/her age group, gender, and occupation.
        \\
        \midrule
        \textbf{Output}:
        \\
        age group:{25-34},gender:{M},occupation:{doctor/health care.}  \\
    \bottomrule
    \end{tabular}
\end{table*}

\begin{table*}[t]
    \centering
        \caption{A data example of the interest recognition task. Data source: Amazon Books Reviews.}
    \label{tab:IR_EXP}
    \small
    \begin{tabular}{p{.95\linewidth}}
    \toprule
        \textbf{Interest Recognition}  \\
    \midrule
        \textbf{Task Description}:
        \\
        The objective of interest recognition is to infer and categorize user interests based on behavioral data, thereby providing interpretable and actionable interest dimensions for personalized recommendation systems.
        \\
        \textbf{Input}:
        \\
        A user has given the following book reviews:
        \\
        Liked books:
        \\
        Little House on the Prairie: I so love these books...I am in my 30s and and just now reading them and they are such a delight...as a child i loved the tv series and the books I must say are even better...
        \\
        Little Lord Fauntleroy: my goodness this is one of the best books ive read....which i am a BIG fan of fhb and have liked everything i have read written by her...this has to be one of my favorites....great story great meaning to the story!!! please read!!!
        \\
        My Sweet Audrina: This was a pretty good read thought I figured out the \"secret\" midway through the book. It's really kind of obvious if you think about it!!! overall thought this was a good book.
        \\
        Disliked books:
        \\
        Silas Marner: The weaver of Raveloe, (The Academy series of English classics): This has to be the most boring book i have ever read! It seems like it takes like 10 pages just to say what could be said in one paragraph. oh my goodness completely boring please dont put your self through this.
        \\
        Can't Wait to Get to Heaven: A Novel: This book was SO disappointing to me. I am a Christin woman and don't like to read accounts of Heaven not taken from the Bible. I first started this book and thought it was going to be great. I stopped on page 134 couldn't go any further. The story was dragged out and overall boring. Waste of time!
        \\
        Silas Marner (New Windmill): This has to be the most boring book i have ever read! It seems like it takes like 10 pages just to say what could be said in one paragraph. oh my goodness completely boring please dont put your self through this.
        \\
        Based on these reviews, briefly summarize the user's reading interests, focusing on the types of books, styles, and authors the user may prefer.
        \\
        \midrule
        \textbf{Output}:
        \\
        The user enjoys classic, wholesome, and sentimental stories (e.g., Little House on the Prairie, Little Lord Fauntleroy) with strong moral themes and emotional depth, possibly favoring older or traditional authors (Frances Hodgson Burnett). They dislike slow-paced, overly verbose classics (Silas Marner) and non-biblical religious fiction (Can't Wait to Get to Heaven). They also appreciate mystery/drama with twists (*My Sweet Audrina) but prefer clear, engaging storytelling without excessive filler.In short: Classic, heartfelt, morally uplifting books with brisk pacing; dislikes tedious or theologically speculative works.  
        \\
    \bottomrule
    \end{tabular}
\end{table*}

\begin{table*}[t]
    \centering
        \caption{A data example of the target user identification task. Data source: Steam.}
    \label{tab:TUI_EXP}
    \small
    \begin{tabular}{p{.95\linewidth}}
    \toprule
        \textbf{Target User Identification}  \\
    \midrule
        \textbf{Task Description}:
        \\
        The objective of target user identification is to accurately pinpoint the user groups most likely to exhibit positive feedback (e.g., clicks, purchases, high ratings) by analyzing the alignment between item characteristics and user profiles, thereby providing targeted distribution guidance for recommendation systems.
        \\
        \textbf{Input}:
        \\
        The attributes of the game Cooking Academy Fire and Knives are as follows: \\
        genres: ['Action', 'Casual'], \\
        tags: ['Casual', 'Action', 'Simulation'], \\
        specs: ['Single-player'], publisher: Viva Media,\\
        Customers' comments on the game are as follows:\\
        1. Just theraputic \& fun to play if you enjoy cooking.\\
        2. 10/10 I am the next Rachel Ray now because this game taught me how to crack an egg properly.\\
        3. Fun and very nice to play. Can be a bit stressful at times though.\\
        Based on the attributes and the reviews, briefly describe which user groups this game is suitable for recommending.
        \\
        \midrule
        \textbf{Output}:
        \\
      Based on the game's attributes and customer reviews, Cooking Academy Fire and Knives is best suited for the following user groups:\\  
      1. Casual Gamers – The game's genres (Action, Casual) and tags (Casual, Simulation) suggest it's easy to pick up and play without requiring deep commitment.\\
      2. Cooking Enthusiasts – Reviews highlight its therapeutic and educational aspects (e.g., learning cooking techniques like cracking eggs), making it appealing to those who enjoy cooking-themed games.\\
      3. Players Seeking Relaxing Yet Engaging Experiences – While mostly fun and therapeutic, the slight stress mentioned in reviews suggests it offers light challenges, fitting players who enjoy mild intensity.\\
      4. Single-Player Gamers – Since it’s single-player only, it suits those who prefer solo gameplay.\\
      Avoid recommending to competitive multiplayer fans or those seeking hardcore action/simulation depth.
        \\
    \bottomrule
    \end{tabular}

\end{table*}
\end{document}